\documentclass[11pt,a4paper]{article}
\usepackage{jheppub}

\usepackage{amssymb,amsbsy,amsfonts,amssymb,amsmath}
\usepackage{natbib,multibib}
\usepackage{graphics,epsfig}
\usepackage{hyperref}
\usepackage{epsdice}

\newcommand{\eq}{\begin{equation}}
\newcommand{\eqe}{\end{equation}}

\newcommand{\eqa}{\begin{eqnarray}}
\newcommand{\eqae}{\end{eqnarray}}

\hypersetup{
    colorlinks=true,       
    linkcolor=red,          
    citecolor=blue,        
    filecolor=magenta,      
    urlcolor=blue           
}

\newcommand{\cO}{\mathcal{O}}
\newcommand{\cS}{\mathcal{S}}

\begin{document}

\preprint{IPMU-15-0224,\,
INT-PUB-15-077}

\title{Spectral sum rules for confining large-$N$ theories}
\author[1]{Aleksey Cherman}
\emailAdd{aleksey.cherman.physics@gmail.com}

\affiliation[1]{Institute for Nuclear Theory, University of Washington, Seattle, USA}%
\author[2,3]{David A.\ McGady}
\emailAdd{mcgady@nbi.ku.dk}
\affiliation[2]{Kavli IPMU, University of Tokyo, Kashiwa, Chiba 277-8586, Japan}
\affiliation[3]{Niels Bohr International Academy \& Discovery Center, Niels Bohr Institute, \\ University of Copenhagen, 2100 Copenhagen, Denmark}
\author[2,4]{and Masahito Yamazaki}
\emailAdd{masahito.yamazaki@ipmu.jp}

\affiliation[4]{Institute for Advanced Study, School of Natural Sciences, Princeton NJ 08540, USA}
\date{\today}

\abstract{
We consider asymptotically-free four-dimensional large-$N$ gauge theories with massive fermionic and bosonic adjoint matter fields, compactified on squashed three-spheres, and examine their regularized large-$N$ confined-phase spectral sums.  The analysis is done in the limit of vanishing 't Hooft coupling, which is justified by taking the size of the compactification manifold to be small compared to the inverse strong scale $\Lambda^{-1}$.  Our results motivate us to conjecture some universal spectral sum rules for these large $N$ gauge theories.
}

\maketitle

\section{Introduction} 
In the large $N$ limit, confining gauge theories become free, in the sense that the interactions of the physical finite-energy degrees of freedom, the mesons and glueballs, become suppressed by positive powers of $1/N$.  But a solution of most such theories has been far out of reach.  To have a chance of solving the large $N$ limit of confining theories, it is essential to understand their symmetry structure.  For instance, before studying connected correlation functions of more than two operators, one would want to understand whether the spectrum is organized by any emergent symmetries at $N = \infty$.   The conjecture that such symmetries might exist has a long history\cite{Polyakov:1975rr,Polyakov:1977vm,Polyakov:1979gp,Polyakov:1980ca}, and has been given explicit support for $\mathcal{N}=4$ super-Yang-Mills theory, see e.g. \cite{Dolan:2003uh,Beisert:2010jr}.  But it has been very difficult to explore the existence and nature of emergent large $N$ symmetries in non-supersymmetric confining theories, because in flat space such theories are strongly coupled at distances which are large compared to the inverse strong scale $\Lambda^{-1}$, and so they are not easily amenable to analytic calculations.  In this paper we gather some evidence for the existence and nature of emergent large $N$ symmetries in such theories by exploring the properties of their confined-phase spectral sums in a tractable limit. 

We will study asymptotically-free 4D large-$N$ with $n_S$ adjoint scalars and $n_F$ adjoint fermions.  The matter fields will be allowed to be massive, and our spacetime geometry will be $\cS^3_R \times \mathbb{R}$, where $\cS^3_R$ is a squashed three-sphere of overall size $R$.  Even though generic large-$N$ theories are free in terms of the interactions between the physical degrees of freedom, in practice we generally only know how to do analytic calculations using the microscopic quark and gluon fields. This guides our choice of geometry, in the following way. Interactions between microscopic fields are characterized by the 't Hooft coupling $\lambda$, and generic non-supersymmetric large-$N$ gauge theories at long distances are strongly coupled in terms of $\lambda$. Working in the regime where $R\Lambda \ll 1$ allows us to avoid this strong coupling problem, because then the 't Hooft coupling at $R$, the longest distance scale in the theory, becomes small: $\lambda = \lambda[1/R] \to 0$.

Consequently, our results will be derived in a doubly weakly-coupled regime where both $1/N$ and $\lambda$ are sent to zero.\footnote{We note that we consider only $SU(N)$ gauge interactions: any other couplings, such as scalar-self-couplings, are set to zero.} This has two utilities. First, in this setting gauge theories are tractable analytically. Secondly, despite being weakly coupled, these systems stay in their confined phase at low temperature, in the sense that they have an unbroken $\mathbb{Z}_N$ center symmetry, and their free energies scale as $N^0$ \cite{Aharony:2003sx}.  For large $N$ and $\lambda = 0$, the large-$N$ confined-phase spectrum of excitations on $\cS^3 \times \mathbb{R}$ --- i.e. the energies and degeneracies of excitations, $\{ \omega_n, d_n\}$, where $n$ is the excitation level number --- can be calculated explicitly.

We should emphasize that we take the limit of large $N$ while holding all other parameters  --- such as the matter content, the strong scale $\Lambda$, matter field masses, and the IR cutoffs (box size parameters) --- fixed.  We also choose to hold the UV cutoff $\mu$ fixed as $N$ is taken to infinity.   This has a very important consequence for our analysis: the fact that $\mu$ is fixed as $N\to \infty$ means that we will only consider states with energies of order $N^0$ throughout this paper.

If large-$N$ emergent symmetries indeed exist, they should produce some interesting features in the large-$N$ confined-phase spectrum. In this paper we gather evidence for two such features, in the form of universal sum rules for confined-phase spectra for non-conformal gauge theories at large-$N$ and $\lambda \to 0$, with multiple IR mass-scales associated with squashing the $\cS^3$ and the masses. Our results significantly generalize our previous work \cite{Basar:2014hda}, written with G.~Ba\c sar, which focused on large-$N$ theories with massless adjoint matter on small round three-spheres, which have only one mass scale.  As is discussed in more detail in the conclusions, our work fits closely into the program on the exploration of emergent symmetries of confining large-$N$ theories pursued in \cite{Basar:2013sza,Basar:2014hda,Basar:2015xda,Basar:2015asd}.

Specifically, we gather evidence for two conjectural large-$N$ sum rules:
\begin{align}
E &= \frac{1}{2}\sum_{n=0}^{\infty}  d_n \omega_n \bigg|_{\rm renormalized} = 0  \; ,
\label{eq:mainResult} \\
\tilde{E} &= \frac{1}{2}\sum_{n=0}^{\infty}  (-1)^Fd_n \omega_n \bigg|_{\rm renormalized} = 0 \; . 
\label{eq:mainResult_without_sign}
\end{align}
Taken at face value, QFT spectral sums such as \eqref{eq:mainResult} and \eqref{eq:mainResult_without_sign} diverge. To make them meaningful one must specify a regularization and renormalization scheme. We choose to regularize the spectral sums using the spectral heat-kernel method, so that
\begin{align}
E &\to E(\mu) := \frac{1}{2}\sum_{n=0}^{\infty}  d_n \omega_n  e^{- \frac{\omega_n}{\mu}} 
=  \frac{1}{2}\sum_{n=0}^{\infty}  d_n \omega_n  q^{\frac{\omega_n}{\omega}} \label{E_mu} \;, \\
\tilde{E} &\to \tilde{E}(\mu) := \frac{1}{2}\sum_{n=0}^{\infty} (-1)^{F}d_n \omega_n  e^{- \frac{\omega_n}{\mu}} 
= \frac{1}{2}\sum_{n=0}^{\infty} (-1)^{F} d_n \omega_n  q^{\frac{\omega_n}{\omega}}  \;, \label{tE_mu}
\end{align}
where $q: = e^{-1/(R\mu)}$, and $1/R = (2\pi^2/\mathrm{Vol}_{\cS^3})^{1/3}$ is a characteristic frequency related to the space geometry, which sets the scale on which the momenta are quantized. For instance, if the spatial manifold is a round 3-sphere, then the parameter $R$ is just its radius.  We again emphasize that, because we take the large-$N$ limit before the $\mu \to \infty$ limit, our spectral sums only include contributions from states with $\omega_n \sim N^0$.  We then define the renormalized value of $E$ and $\tilde{E}$ by identifying them with the finite terms $E_0$ and $\tilde{E}_0$ in the $\mu \to \infty$ expansion of $E(\mu)$ and $\tilde{E}(\mu)$:
\begin{align}
E(\mu) &= E_4 \mu^4 R^{3} + E_2 \mu^2 R + E_0 + \cO\left(1/\mu \right)  \;, \label{eq:E_mu_expand} \\
\tilde{E}(\mu) &= \tilde{E}_4 \mu^4 R^3+ \tilde{E}_2 \mu^2 R + \tilde{E}_0 + \cO\left(1/\mu \right) \label{eq:tE_mu_expand} \,.
\end{align}
We hasten to add that the fact that the large $\mu$ expansion takes the form above is not meant to be obvious: it is actually a non-trivial part of our results.   A priori, one could imagine that the large $\mu$ expansion of a heat-kernel-regulated QFT spectral sum would  include a term proportional to $\log(\mu R)$, and indeed in a generic QFT with massive fields such terms do arise.  When this happens, it does not make sense to define renormalized values for $E$ and $\tilde{E}$ as the finite parts of $E(\mu)$ and $\tilde{E}(\mu)$, because the finite parts become badly scheme-dependent, in the sense that linear rescalings of $\mu$ shift the finite parts, $E_0$ and $\tilde{E}_0$. However, in all of the confining large-$N$ gauge theory examples we have explored, $\log(\mu R)$ terms are actually power suppressed, appearing as $\mu^{-p} \log(\mu R)$ with $p >0$.  As a result, the large-$\mu$ expansion coefficients  $E_{i}$ and $\tilde{E}_i$ shown above are not ambiguous. We find evidence supporting the conjecture that 
\begin{align} 
E_0 = 0 \;, \qquad \tilde{E}_0 = 0 \;, \label{eq2vanish} \\
E_4 = 0 \;, \qquad \tilde{E}_4 = 0 \, , \label{eq2D_Hint}
\end{align}
for all large-$N$ gauge theories on $\cS^3 \times \mathbb{R}$, in the limit $\lambda=0$ of fermionic and bosonic adjoint matter fields, with arbitrary masses. 

The plan of this paper is as follows. In Section~\ref{sec.general}, we explain the structure of the spectral sums, recalling their relations to the single-trace partition functions for large-$N$ gauge theories, and motivate the spectral sum rules. In Section~\ref{sec:leading}, we explain the argument motivating the conjectured sum rules in~\eqref{eq2vanish} and~\eqref{eq2D_Hint}. In Section~\ref{sec:examples} we then explicitly work out the leading-order effects of turning on mass terms for the matter fields and deforming $\cS^3$ away from the round-sphere limit.  The results are consistent with the conjectured sum rules.  We comment on the interpretation of our findings and conclude in Section~\ref{sec:conclusion}.

\section{Setting up the calculation}
\label{sec.general}

In this section, we set up the mathematical framework for evaluating~\eqref{eq:mainResult} and~\eqref{eq:mainResult_without_sign} in the theories described in the introduction.   Our goal is to evaluate \eqref{E_mu} and \eqref{tE_mu}, where $d_n$ and $\omega_n$ are the degeneracies and energies of the single-particle excitations of the confined phase of our large-$N$ theories. To do this, observe that, if we write $q = e^{-1/(R\mu)} = e^{-\beta/R}$, then
\begin{align}
E(\mu) &= \frac{1}{2}\frac{q}{R} \frac{d}{dq }\sum_{n}d_n q^{R \omega_n}
= \frac{1}{2}\frac{q}{R} \frac{d}{dq} Z_{\rm ST}(\beta) \bigg|_{\beta = 1/\mu}  \;,\\
\tilde{E}(\mu) &= \frac{1}{2}\frac{q}{R} \frac{d}{dq }\sum_{n}  (-1)^F d_n q^{R \omega_n}
= \frac{1}{2}\frac{q}{R} \frac{d}{dq} \tilde{Z}_{\rm ST}(\beta) \bigg|_{\beta = 1/\mu}  \;,
\label{eq:twistedSum}
\end{align}
where 
\begin{align}
Z_{\rm ST}(\beta) &= \sum_{n\ge 0}  d_n \, q^{R\omega_n } =\sum_{n\ge 0}  d_n \,  e^{-\frac{\omega_n}{\mu} }\, ,\\
\tilde{Z}_{\rm ST}(\beta) &= \sum_{n\ge 0} (-1)^F d_n\, q^{R\omega_n } =\sum_{n\ge 0} (-1)^F d_n \, e^{-\frac{\omega_n}{\mu} } .
\end{align}  
Indeed, these last expressions are precisely the single-particle thermal and $(-1)^F$-twisted confined-phase partition functions. At large $N$, the confined-phase single-particle partition functions are precisely single-trace partition functions, motivating the notation. 

It is possible to write down fairly explicit expressions for $Z_{\rm ST}$ and $\tilde{Z}_{\rm ST}$ for the class of QFTs we consider.  In the $\lambda \to 0$ limit, these systems can be thought of as collections of adjoint harmonic oscillators with a Gauss-law constraint that forces the physical states to be color singlets~\cite{Aharony:2003sx,Polyakov:2001af,Sundborg:1999ue}. So to write down the single-trace partition functions, it is first useful to determine the `harmonic oscillator' partition functions associated with the elementary free gluon and adjoint-matter fields (fermionic or bosonic), which we respectively denote by $z_V(q;\vec{\epsilon})$ and $z_{F,S}(q; \vec{m}, \vec{\epsilon})$, where $\vec{\epsilon}$ are squashing parameters, while $\vec{m}$ are mass parameters.  These `single-letter' partition functions explicitly depend on the mass parameters $\vec{m}$ as well as the deformation parameters $\vec{\epsilon}$ of the three-sphere, and encode the single-particle excitation energies and degeneracies of free vector, scalar, and Majorana fermion fields on $\cS^3 \times \mathbb{R}$.   In terms of $z_{V,F,S}$, the single-trace partition functions take the form~\cite{Aharony:2003sx}
\begin{align}
Z_{\rm ST}(\beta; \vec{m},\vec{\epsilon})& =- \sum_{n=1}^{\infty} \frac{\varphi(n)}{n} \log \left[
1-z_V(q^n;\vec{\epsilon}) -n_S z_S(q^n; \vec{m}, \vec{\epsilon})+ (-1)^n n_F z_F (q^n; \vec{m}, \vec{\epsilon}) \right] \;,
\label{eq:thermalZ}\\
\tilde{Z}_{\rm ST}(\beta; \vec{m},\vec{\epsilon})& = -\sum_{n=1}^{\infty} \frac{\varphi(n)}{n} \log \left[ 
1-z_V(q^n;\vec{\epsilon}) -n_S z_S(q^n; \vec{m}, \vec{\epsilon}) + n_F z_F (q^n; \vec{m},\vec{\epsilon}) \right] \;,
 \label{eq:twistedZ} 
\end{align}
where $\varphi(k)$ is the Euler totient function, and all of the dependence on the mass parameters $\vec{m}$ and the geometry of $\cS^3$ contained in $\vec{\epsilon}$ enters through the single-letter partition functions. We will give explicit expressions for the single-letter partition functions, $z_V(q^k;\vec{\epsilon})$, $z_S(q^k; \vec{m}, \vec{\epsilon})$, and $z_F (q^k; \vec{m},\vec{\epsilon})$ below.

\section{Motivation for sum rules}
\label{sec:leading}

In this section we explain a slightly naive approach to the computation of $E_0, \tilde{E}_0$ and $E_4, \tilde{E}_4$, which gives results that  motivate our conjectured sum rules \eqref{eq2vanish} and \eqref{eq2D_Hint}.  As will become clear, the calculation in Section~\ref{sec:E0} drops some potential contributions to $E(\mu)$ because they appear to be negligible in the large $\mu$ limit.  The reason we call the approach naive is that these contributions do, in fact, contribute to the large $\mu$ asymptotics, as we will explain in \ref{sec:subleading}.    However, in every example we have been able to check explicitly, the neglected effects only contribute to the coefficient of $\mu^2$ at large $\mu$, and do not affect the coefficients of $\mu^0$ or $\mu^4$, supporting the conclusions drawn from the simple calculations given in this section.

\subsection{Asymptotics of partition functions and the sum rules}
\label{sec:E00}
To understand the behavior of the spectral sums for large $\mu$, we must understand the behavior of the single-trace partition functions as $q$ approaches $1$ from within the unit disk.  As discussed in \cite{Basar:2014hda}, the limit $q\to 1$ ($\mu\to \infty$) should not be taken along the real axis of $q$, since depending on the fermion boundary conditions and masses there can be poles on the real $q$ axis, corresponding to Hagedorn instabilities of large-$N$ confining theories. Instead, we take the $q\to 1$ limit by approaching the point $q=1$, from the inside of the unit disc $|q|<1$ in the complex $q$-plane, along a path that does not go through Hagedorn poles.

In view of the structure of the single-trace partition functions, their $q \to 1$ behavior is dictated by the $q \to 1$ behavior of the single-letter partition functions. Expanding the single-letter partition functions about $q = 1$ yields,  
\begin{align}
\begin{split}
z_V(q; \vec{\epsilon}) &\to \frac{v_{-3}}{(1-q)^{3}}+\frac{v_{-2}}{(1-q)^{2}} + \frac{v_{-1}}{(1-q)}  +  \cO\left[ (1-q)^{0}\right] \;, \\
z_F(q; \vec{m}, \vec{\epsilon}) &\to \frac{f_{-3}}{(1-q)^{3}} +\frac{f_{-2}}{(1-q)^{2}}  + \frac{f_{-1}}{(1-q)}   +  \cO\left[ (1-q)^{0}\right] \;, \\
z_S(q; \vec{m}, \vec{\epsilon}) &\to \frac{s_{-3}}{(1-q)^{3}}+\frac{s_{-2}}{(1-q)^{2}} + \frac{s_{-1}}{(1-q)}  +  \cO\left[ (1-q)^{0}\right]\;, \label{eq:SingleLetterAsymptotics}
\end{split}
\end{align}
where a priori one would expect $v_{i}, s_{i}, f_{i}$ to depend on $\vec{m}$ and $\vec{\epsilon}$. A key point of this section is that, on very general grounds, there is a simple relation between the two leading Laurent-series coefficients within~\eqref{eq:SingleLetterAsymptotics}:
\begin{align}
\frac{v_{-2}}{v_{-3}}=\frac{f_{-2}}{f_{-3}} =\frac{s_{-2}}{s_{-3}}  = -\frac{3}{2}\;. \label{eq:AsymptoticsConstraints}
\end{align}

This result can be understood as follows.  First, we note that because $q \to 1$ is a high-energy limit, one can conclude that $v_{-3}, f_{-3}$ and $s_{-3}$ are determined by the physics of free-vector, free-fermion, and free-scalar QFTs in flat space.  Indeed, these coefficients control the coefficients of the leading $T^4$ term in the free energy of these QFTs, so that for instance  $F_{\rm scalar} = \frac{\pi^2}{90} s_{-3} \mathrm{Vol}_{\cS^3} T^4$.  This means that $v_{-3}, f_{-3}, s_{-3}$ are non-vanishing.  Moreover, they must be \emph{independent} of any mass or squashing parameters, and so are fixed by the number of degrees of freedom to be
\begin{align}
v_{-3} = 4\;, \quad s_{-3} = 2\;, \quad f_{-3} = 4 \;. \label{eq:leadingCoeffs}
\end{align}

Second, we show that the coefficients of $(1-q)^{-3}$ determine the coefficients of $(1-q)^{-2}$, within a single-letter partition function, such that~\eqref{eq:AsymptoticsConstraints} holds. To see this, it is helpful to study the spectral sum for a single free Majorana fermion on $\cS^3_R \times \mathbb{R}$, with canonical partition function given by $z_F(q; \vec{m}, \vec{\epsilon})$ in~\eqref{eq:SingleLetterAsymptotics}. The regularized spectral sum $E_F(\mu)$ is
\begin{align}
E_F(\mu)&=  -\frac{1}{2} q \frac{d}{dq}z_F(q) \nonumber \\ 
&= \frac{1}{2} \left[3 f_{-3}\mu^4 R^3+ (3 f_{-3} +  2f_{-2}) \mu^3 R^2   + ( f_{-3} + f_{-2} + f_{-1})\mu^2  R + \mathcal{O}(\mu^0) \right].
\end{align}
The values of $f_{-3},f_{-2},f_{-1}$ are determined by the choice of a specific manifold $\cS^3$ and the fermion mass. As we will now explain, a $\mu^3$ divergence is forbidden in a Poincar$\acute{\textrm{e}}$-invariant field theory, and this implies $f_{-3}/f_{-2} = -2/3$.

To understand why $\mu^3$ divergences are forbidden, we classify the possible generally-covariant counter-terms.  The only counter-term with mass dimension $4$ is $\mu^4 \int d^{4}x \sqrt{g}$.  This means that the $\mu^4$ divergence is (a) possible, so that in general one should expect $f_{-3} \neq 0$, and (b)  it can be absorbed by adjusting the coefficient of $\mu^4 \int d^{4}x \sqrt{g}$ counter-term. Similarly, since one can write the counter-terms $\mu^2 \int d^{4}x \sqrt{g} \, \mathcal{R}$ and $\mu \int d^{4}x \sqrt{g} \, \bar{\psi}\psi$, one can expect $f_{-3}+f_{-2} + f_{-1}$ to be non-zero in general. But there are no generally-covariant counterterms of dimension $1$ in four spacetime dimensions, so there cannot be a $\mu^3$-divergence.  The same arguments go through for the spectral sum for a massless vector field. In theories with scalars, this conclusion requires imposing a global $\mathbb{Z}_2$ $\phi \to -\phi$ symmetry on the scalar. But in gauge theories, which is the setting we are interested in, such a $\mathbb{Z}_2$ symmetry is always automatically present as a consequence of gauge invariance. 

Thus we learn that $3 f_{-3} +  2f_{-2}$, $3 s_{-3} +  2s_{-2}$, and $3 v_{-3} +  2v_{-2}$ must all vanish in all relevant consistent Poincar$\acute{\textrm{e}}$-invariant QFTs. As a result, the first two coefficients in the Laurent expansion of the single-letter partition functions are independent of mass parameters or deformations of the space geometry. Of course, the coefficients of $(1-q)^{k}$ for $k>-2$ do depend on mass parameters and geometric deformations, and in general the $q\to1$ expansion includes terms which are non-analytic in $(1-q^n)$ beyond the order which we considered above. We revisit the contributions of the $(1-q)^{k}, k>-2$ terms in section~\ref{sec:subleading}.

\subsection{Sum rule for $E_0$}
\label{sec:E0}
We now want to evaluate the large $\mu$ limit of the spectral sums.  The relations between the spectral sums and the single-trace partition functions suggest that we should explore the behavior of 
\begin{align}
q \frac{d}{dq} Z_{\rm ST} &=  -\sum_{n=1}^{\infty} \varphi(n) q^n \frac{d}{d q^n}\log\left[1-z_V(q^n)-n_S z_S(q^n) +(-1)^n n_F z_F (q^n) \right] \;,  \\
q \frac{d}{dq} \tilde{Z}_{\rm ST} &=  -\sum_{n=1}^{\infty} \varphi(n) q^n \frac{d}{d q^n}\log\left[1-z_V(q^n)-n_S z_S(q^n) + n_F z_F (q^n) \right] \;.
\label{eq:sum}
\end{align}
To evaluate the asymptotics of the functions in \eqref{eq:sum}, we observe that generically --- i.e., in the absence of a $(-1)^F$-twist for supersymmetric matter content, $n_S = 2( n_F - 1)$ --- the coefficient of $(1-q)^{-3}$ in the $q \to 1$ expansion of the argument of the logarithms above is non-vanishing. (We will treat the case of $n_S = 2 n_F - 2$ in section~\ref{sec:SUSY}.)  In the generic situation, we find that
\begin{align}
&q^n\frac{d}{dq^n} \log \left[ 1-z_V(q^n)-n_S z_S(q^n) +c_n\, n_F z_F (q^n)  \right] \nonumber\\
&\qquad = \frac{3q^n}{1-q^n} - \frac{ n_S s_{-2}  +c_n\, n_F f_{-2}    - v_{-2}}{  n_Ss_{-3} +c_n\,  n_F f_{-3}   - v_{-3}} q^n +\cO(1-q^n) \;, \label{c1c0}
\end{align}
where $c_n = (-1)^n$ for $Z_{\rm ST}$ and $c_n=1$ for $\tilde{Z}_{\rm ST}$.  The two terms which are explicitly shown above are both non-vanishing as $q^n \to 1$.  At this stage, it is important to observe that these non-vanishing $(1-q^n)^{-1}$ and $(1-q^n)^0$ terms in \eqref{c1c0} depend only on leading asymptotics of the single-letter partition function, and are manifestly independent of microscopic mass parameters and squashing parameters. The terms which are not explicitly written above depend on the non-universal --- and in general non-analytic in $(1-q^n)$ --- subleading asymptotics of the single-letter partition functions, and are suppressed by positive powers of $(1-q^n)$. Because of this suppression, they vanish as $q^n \to 1$ for fixed $n$, and we neglect them in this section. We return to them in Sections~\ref{sec:subleading} and \ref{sec:mass_terms}.

The coefficient $3$ multiplying $(1-q^n)^{-1}$ is actually $d-1$ (with $d$ the spacetime dimension), and comes from logarithmic derivatives of the $(1-q)^{-3} \sim \mu^3$ terms in the single-letter partition functions.  Naively the coefficient of $(1-q^n)^0$ in~\eqref{c1c0} depends on $n_F, n_S$ and the parameter $c_n$, which reflects whether we insert $(-1)^F$ into the spectral sum. However, by~\eqref{eq:AsymptoticsConstraints}, all of this dependence cancels out, and 
\begin{align}
&q^n\frac{d}{dq^n} \log \left[ 1-z_V(q^n)-n_S z_S(q^n) \pm n_F z_F (q^n) \right] \nonumber\\
&\qquad= \frac{3q^n}{1-q^n} + \frac{3}{2}q^n+ \mathcal{O}(1-q^n) \;. \label{Fc1c0}
\end{align}
To compute $E_0$ and $E_4$, we simply sum over $n$ in \eqref{eq:sum}, and take the limit $q \to 1$ with the assumption that the subleading terms in \eqref{c1c0} do not contribute to the non-vanishing parts of the spectral sum in this limit.  Then the computation of the spectral sum reduces to understanding the behavior of
\begin{align}
E(\mu) = \tilde{E}(\mu) =-\frac{1}{2 R}\sum_{n=1}^{\infty} \varphi(n) \left[  \frac{3q^n}{1-q^n} + \frac{3}{2}q^n \right] \,.
\end{align}
To evaluate this expression, let us define the function
\begin{align}
f_{m}(\mu)\bigg|_{q = e^{-1/\mu}} &= \sum_{n=1}^{\infty} \varphi(n)\, q^n(1-q^n)^m .
\label{eq:f_m_definition}
\end{align}
For our immediate purposes we only need the large-$\mu$ asymptotics of $f_{0}$ and $f_{-1}$, which can be shown to be
\begin{align}
f_{0}(\mu)\bigg|_{q = e^{-1/\mu}} &= \sum_{n=1}^{\infty} \varphi(n) \, q^n = \frac{6}{\pi^2} (\mu R)^2+ \frac{1}{6} + \mathcal{O}\left(\frac{1}{\mu R}, \frac{1}{\mu R}\log(\mu R)\right) \;, \label{eqF0} \\
f_{-1}(\mu)\bigg|_{q = e^{-1/\mu}} &= \sum_{n=1}^{\infty} \varphi(n)\, \frac{q^n}{1-q^n} = (\mu R)^2-\frac{1}{12}  + \mathcal{O}\left(\frac{1}{\mu R}, \frac{1}{\mu R}\log(\mu R)\right)  \;,\label{eqFM1} 
\end{align}
and the coefficients of the $1/\mu^k, k>0$ terms in $f_{-1}(\mu)$ can be shown to be $\zeta(-k)$ times the coefficients of the $1/\mu^k$ terms of $f_{0}(\mu)$.  Noting that neither function includes a $\mu^4$ term, we conclude that $E_4$ and $\tilde{E}_4$ vanish.  Both functions have a non-vanishing $\mu^0$ term.  But in the combination of $f_{0}$ and $f_{-1}$ relevant for our spectral sum, the constant terms cancel, allowing us to conclude that $E_0 = \tilde{E}_0 = 0$ as well.

\subsection{Sum rule for $E_4$}  
\label{sec:E4}
The vanishing of $E_4$ and $\tilde{E}_4$ can be understood from existing results in the literature.  The key point is that the coefficient of $\mu^4$ in a spectral sum is controlled by the high-energy properties of the spectrum.  But as far as very high-energy states are concerned, the mass parameters and the geometry of the compactification manifold are irrelevant. The properties of the high-energy states are controlled by the UV fixed point of the theory, so that the value of the coefficient of $\mu^4$ can be determined from the behavior of free massless large-$N$ gauge theories in flat space.  The spectrum of local operators of such theories is encoded in their round-$S^3$ partition functions. It was recently shown that the grand canonical confined-phase partition functions for $\lambda = 0$ large-$N$ gauge theories with massless adjoint matter, with and without insertions of $(-1)^F$ are (vector-valued, meromorphic) modular forms, with well-defined modular weight~\cite{Basar:2014jua,Basar:2015xda} (see also~\cite{Zuo:2015mxk}). Because these partition functions are modular, it can be shown that at small-$\beta$ they must scale $e^{\sigma/\beta}$ for some number $\sigma$.  (Here by  `small-$\beta$' we mean  the limit where $|\beta|$ is taken to zero before  $\arg \beta$ is taken to zero, in order to avoid Hagedorn instabilities which might be present along the ray $\arg \beta = 0$.  For more on this issue see \cite{Basar:2014hda,Basar:2015xda,Basar:2015asd}.)  Equivalently, $Z(\mu) \sim e^{\sigma \mu}$ as $\mu \to \infty$.  Further, because $Z$ and $Z_{\rm ST}$ (and also $\tilde{Z}$ and $\tilde{Z}_{\rm ST}$) are related by the plethystic exponential,
\begin{align}
Z = e^{-\sum_{n\ge1} \frac{1}{n} Z_{\rm ST}(n\beta)} \, ,
\end{align}
we may infer that $Z_{\rm ST}(\beta) \sim \beta^{-1}$, \emph{not} $\sim \beta^{-3}$. We show this by contradiction. Suppose that $Z_{\rm ST}(\beta \to 0) \sim \beta^{-3}$.  Then since $\sum_{n\ge1} n^{-4} = \pi^4/90 \neq 0$, we would find that $\log Z \sim 1/\beta^3$. However, this contradicts the known modularity properties of $Z$, which imply that $\log Z \sim 1/\beta$, so we learn that $E_4$ and $\tilde{E}_4$ must vanish for the class of theories we are considering.

\subsection{Value of $E_2$}
\label{sec:subleading}
In deriving  the large $\mu$ ($q \to 1$) asymptotics of the spectral sums we worked in two steps, motivated by the fact that the single-trace partition functions $Z(q)$ are built from infinite sums in $n$ of functions of $q^n$, see \eqref{eq:thermalZ} and \eqref{eq:twistedZ}. First, we worked at fixed $n$ and extracted the terms that are non-vanishing as $q^n \to 1$.  Second, taking these non-vanishing terms, we summed over $n$ to obtain the behavior of $Z(q)$ for $q \to 1$.  

In this section, we discuss the effect of the terms which were neglected in the computation in Section~\ref{sec:E0} because they vanished as $q^n \to 1$. We do so in the simplest example, which is large-$N$ confining theories with massless adjoint matter on a round three-sphere.  These terms did not enter the zeta-function-regularization calculations in \cite{Basar:2014hda}, but were automatically taken into account in the two alternative methods of calculations in \cite{Basar:2014hda} using temperature-reflection symmetry \cite{Basar:2014mha} and the direct numerical evaluation of the spectral sums done in \cite{Basar:2014hda}.  Consequently, the neglected terms do not contribute to the coefficient of $\mu^0$ in the large $\mu^0$ expansion of the spectral sums.  (This was not explicitly discussed in  \cite{Basar:2014hda}.)   Here we examine these terms directly, and verify that they do not  contribute to the coefficients of $\mu^4$ or $\mu^0$ in the large $\mu$ expansion.  This point was also made in \cite{Zuo:2015mxk}.  Nevertheless, the neglected terms do contribute to the coefficient of $\mu^2$, which is in fact necessary for consistency with the results of \cite{Basar:2015xda}.  In the next section, which addresses examples of deformations by mass terms or squashing of the $S^3$, we show an identical set of conclusions also holds in these more general cases.

To explore the contributions of the terms that vanish for fixed $n$ as $q^n \to 1$, we first observe that, for theories with massless matter on a round $S^3$, the linear combination of single-letter partition functions that is relevant for the spectral sum, for instance $1-z_V(q^n) + n_F z_F(q^n) -n_S z_S(q^n)$, is a rational function of $Q = q^{1/2}$.  Consequently, the expansion of $\frac{d}{d Q} \log\left[1-z_V(Q)+ n_F z_F(Q) -n_S z_S(Q)\right]$  near $Q=1$ takes the form of a Laurent series in $1-Q$,
\begin{align}
\tilde{E}(\mu) = \frac{1}{R}Q \frac{d}{dQ} \tilde{Z}_{\rm ST} &= -\frac{1}{R}\sum_n \varphi(n)   Q^n \frac{d}{d Q^n} \log\left[1-z_V(q^n) + n_F z_F(q^n) -n_S z_S(q^n)\right] \nonumber\\
&=\frac{1}{2R}\sum_{n\ge1}\varphi(n)  Q^n \sum_{m\ge-1} c_m (1-Q^n)^{m}\nonumber\\
&=\frac{1}{2R} \sum_{m\ge-1} c_m  f_m(2\mu) \, ,
\label{eq:fullSeriesModular}
\end{align}
where the function $f_{m}(\mu)$ is defined in \eqref{eq:f_m_definition}.  As an example, in the particularly simple case of $\mathcal{N}=4$ SYM, with a $(-1)^F$ twist, we find 
\begin{align}
Q \frac{d}{d Q} \log\left[ 1-z_V(Q)+ 4 z_F(Q) -6 z_S(Q) \right] &= Q \frac{d}{d Q} \log\left[ \frac{(1-Q)(1+Q\, z)(1+Q/z)}{(1+Q)^3} \right] \nonumber\\
&= \frac{-Q}{1-Q} - 3 \cdot \frac{Q}{1+Q} + \frac{z Q}{1 + z Q} + \frac{Q/z}{1+Q/z} \nonumber \\
& = \sum_{\alpha} \frac{p(\alpha) \alpha Q}{1+\alpha Q} \;,
\label{eq:N4rootsum}
\end{align}
where $z: = 2 + \sqrt{3}$, $\sum_{\alpha}$ runs over the singular points --- zeros and poles --- of $1-z_V(Q)+ 4 z_F(Q) -6 z_S(Q)$, and $p(\alpha)$ is the order of the `pole' $\alpha$, such that zeros of order $+m$ are counted as poles of order $-m$.   The Laurent series expansion of $E(\mu)$, expanded about the point $\mu \to \infty$, is found in three steps.  

First, we expand each factor $1/(a+Q)$ about $Q = 1$ to find simple closed-form expressions for the coefficients $c_m$ in Eq.~\eqref{eq:f_m_definition}:
\begin{align}
(m ~\geq~ 0) ~:~& c_m =  3 \left(  \frac{1}{1+1} \right)^{m+1} - \left( \frac{1/z}{1+1/z} \right)^{m+1} - \left( \frac{z}{1+z} \right)^{m+1}
\label{eq:cMexplicit} ~~ {\rm and} \\
(m = -1) ~:~& c_{-1} = -1 \;. \nonumber
\end{align}

Second, we use the binomial expansion to relate the Laurent coefficients of $f_{m > 0}(\mu)$ to those of $f_0(\mu)$, 
\begin{align}
f_m(\mu) &= \sum_{n = 0}^{\infty} \varphi(n) Q^n (1-Q^n)^m = \sum_{n = 0}^{\infty} \varphi(n) Q^n \left( \sum_{k = 0}^{m} \begin{pmatrix} m \\ k \end{pmatrix} (-1)^k Q^{n k} \right) \nonumber\\
&= \sum_{k=0}^{m}  \begin{pmatrix} m \\ k \end{pmatrix} (-1)^k f_0\left(\frac{\mu}{k+1}\right) \nonumber\\
&= \frac{6(\mu R)^2}{\pi^2} \sum_{k = 0}^{m} \begin{pmatrix} m \\ k \end{pmatrix} \frac{(-1)^k}{(k+1)^2} + \frac{1}{6} \sum_{k = 0}^{m} \begin{pmatrix} m \\ k \end{pmatrix} \frac{(-1)^k}{(k+1)^0}    + \mathcal{O}\left(\frac{1}{\mu R}\right) \, .
\label{eq:fMscaling}
\end{align}

\begin{figure}[h!]
\centering
\includegraphics[width=0.7\textwidth]{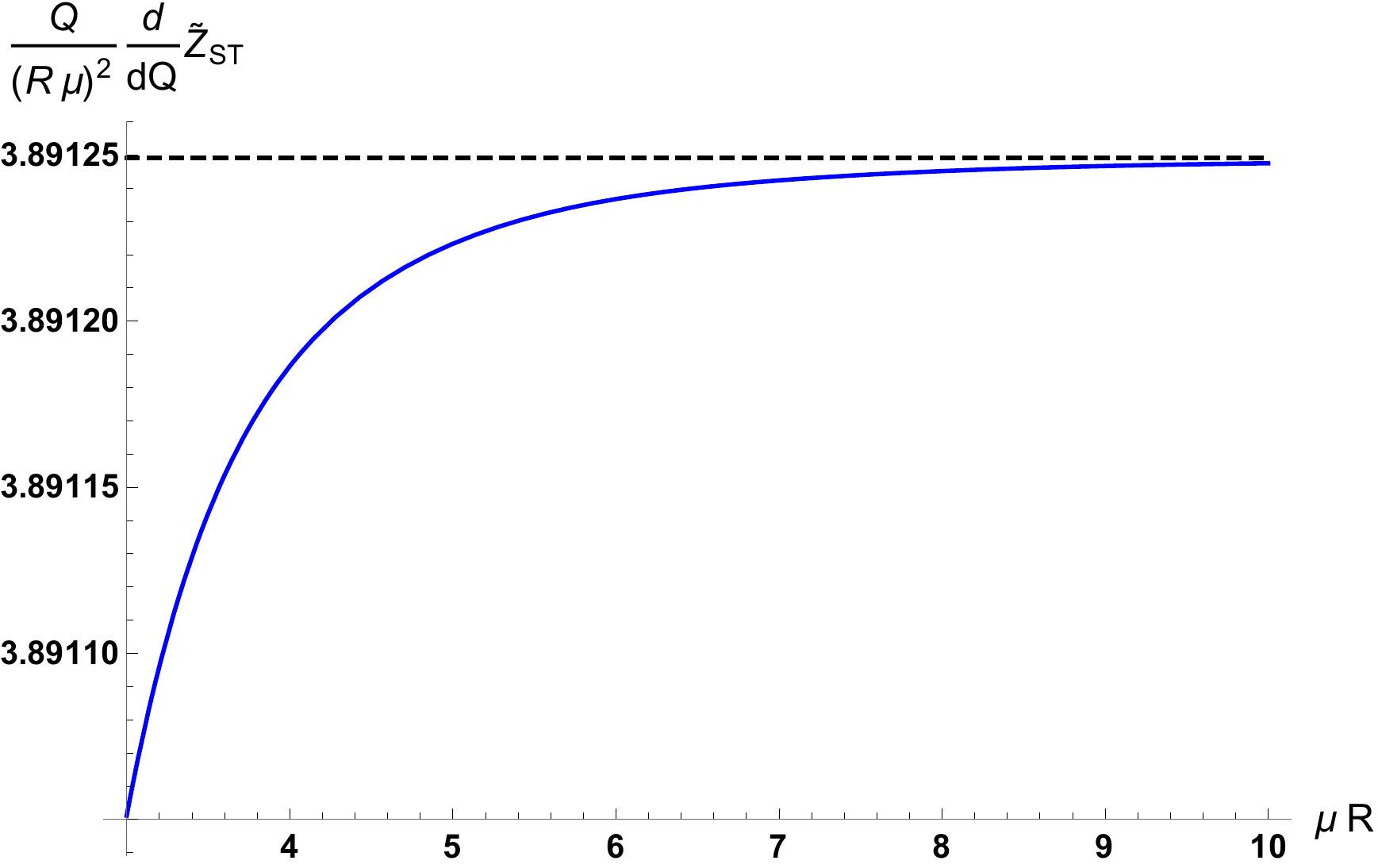}
\caption{(Color Online.) Large $\mu$ behavior of $\frac{Q}{(\mu R)^2}\frac{d}{dQ}\tilde{Z}_{\rm ST}(Q)$ (blue curve) for large-$N$ $\mathcal{N}=4$ SYM theory with massless matter on the round three-sphere, compared to the analytic result for the asymptotic value of the $\mu^2$ coefficient in \eqref{eq:mu2coeffN4} (dashed black line).
}
\label{fig:mu2coeff}
\end{figure}

Third, after putting together \eqref{eq:cMexplicit} with \eqref{eq:fMscaling}, we observe that the general form of the coefficients of $\mu^k$ in Eq.~\eqref{eq:fullSeriesModular} matches a series representation of the polylogarithm function proved in Theorem 2.1 of \cite{Guillera:2008aa}
\begin{align}
\operatorname {Li} _{s}(z)=\sum _{k=0}^{\infty }\left({-z \over 1-z}\right)^{k+1}~\sum _{j=0}^{k}(-1)^{j+1}{k \choose j}(j+1)^{-s}\,.
\end{align} 

Combining these results, we see that the logarithmic $Q$-derivative of twisted ${\cal N} = 4$ SYM's canonical partition function is
\begin{align}
Q \frac{d}{dQ} \tilde{Z}_{\rm ST}^{{\cal N} =4 } &=  \frac{4 \cdot6}{\pi^2} (\mu R)^2 \bigg\{ 
\textrm{Li}_2\left(+1\right) +\textrm{Li}_2\left(-\frac{1}{z}\right) +\textrm{Li}_2\left( -z \right) -3 \textrm{Li}_2(-1) 
\bigg\} + \mu^0 \cdot 0 +  {\cal O}\left( \frac{1}{\mu R} \right) \nonumber\\
&\approx 3.891 (\mu R)^2 + {\cal O}\left( \frac{1}{\mu R} \right) \;, \label{eq:mu2coeffN4}
\end{align}
where polylogarithms of negative weight are simply rational functions of their arguments.  This result agrees with a direct numerical evaluation of $\frac{d}{dQ}\tilde{Z}_{\rm ST}^{{\cal N} = 4}(Q)$ near $Q = 1$, as illustrated in Fig.~\ref{fig:mu2coeff}.  Further, standard dilogarithm identities imply that Eq.~\eqref{eq:mu2coeffN4} agrees with Eq.~(4.12) of~\cite{Basar:2015asd}, which evaluated the same quantity using the modularity of the grand canonical partition function. Equation~\eqref{eq:mu2coeffN4}
is a special case of a more general identity for large-$N$ gauge theories on $S^3 \times S^1$ with arbitrary numbers of massless adjoint scalars and fermions. With the more general function $1-z_V(q)-n_S z_S(q)+ n_F z_F(q)$, we have:
\begin{align}
Q\frac{d}{dQ}\tilde{Z}_{\rm ST}(Q)=\frac{24}{ \pi^2}\left( \sum_{\alpha}p(\alpha) \textrm{Li}_2(-\alpha)\right) (\mu R)^2 + (\mu R)^0 \cdot 0 + \mathcal{O}\left(\frac{1}{\mu R}, \frac{\log(\mu R)}{\mu R}\right)  \, ,
\end{align}
where $\sum_{\alpha}$ runs over the poles and roots of $1-z_V(q)-n_S z_S(q)+ n_F z_F(q)$, and $p(\alpha)$ tracks the order of the associated pole/root. The asymptotics of $Q\frac{d}{dQ}Z_{\rm ST}$ are analogous.

\section{Examples}
\label{sec:examples}

In this section we show calculations supporting our sum rule conjectures in theories with explicitly broken scale invariance.

\subsection{Mass deformation}
\label{sec:mass_terms}

We start by considering the effects of turning on mass terms for the matter fields on the spectral sum in large-$N$ gauge theories.  The single-letter partition function for fermions on a round $S^3$, with mass $m_F$ included, is given by
\begin{align}
z_F(q,M_F) = \sum_{n = 0}^{\infty} 2n(n+1) q^{\sqrt{\left(n+\frac{1}{2}\right)^2 +M_F^2}} \ ,
\end{align}
where we defined the dimensionless parameter $M_F:=m_F R$.  Given $z_F(q,M_F)$, one can in principle compute the heat-kernel-regularized spectral sum of a single fermion on a three-sphere.  In practice, however, we do not know of a useful closed-form expression for $z_F(q,M_F)$.  Nevertheless, it is not hard to work out the form of $z_F(q,M_F)$ order by order in a small $M_F$ expansion. The first two terms are
\begin{align}
z_F(q,M_F) = \frac{4 q^{\frac{3}{2}}}{(1-q)^3} + \frac{M_F^2\log(q)\left[q^{\frac{1}{2}} \left(q+1\right) -(q-1)^2  \tanh ^{-1}\left(q^{\frac{1}{2}}\right)\right]}{2 (q-1)^2} + \mathcal{O}(M_F^4) \;.
\end{align} 
Similarly, for conformally-coupled scalars with a mass $m_S$, we get
\begin{align}
z_S(q,M_S) = \sum_{n = 0}^{\infty} n^2 q^{\sqrt{n^2 +M_S^2}} \ ,
\end{align}
where $M_S := m_S R$, and 
\begin{align}
\begin{split}
z_S(q,M_S) &= \frac{q (q+1)}{(1-q)^3} +\frac{M_S^2 q \log (q)}{2 (q-1)^2} \\
&+ \frac{M_S^4 \log (q) \left[(q-1) \log (1-q)-q \log (q)\right]}{8 (q-1)} +  \mathcal{O}(M_S^6) \;.
\end{split}
\end{align}

It is instructive to look at the expressions for the spectral sum for e.g.\ the massive fermion field.  In terms of $q$, we find
\begin{align}
\begin{split}
\frac{1}{2}q\frac{d z_{F}}{dq} &= E_F(\mu)R = \frac{6}{(1-q)^4} - \frac{12}{(1-q)^3}  + \frac{27-2 M_F^2}{4 (1-q)^2}+ \frac{2 M_F^2-3}{4 (1-q)} \\
&+ \frac{1}{192} \left(24 M_F^2 \log (1-q)+4 M_F^2 \left[5-\log (4096)\right]-9\right)  + \mathcal{O}(1-q) \;.
\end{split}
\end{align}
We note the relation between the coefficients of the first two terms agrees with the general arguments given in Section~\ref{sec:E00}.  It is also to instructive to write the result in terms of $\mu$:
\begin{align}
\label{eq:freeFermionMassive}
E_F(\mu)&=-6 R^3\mu^4 + \frac{1}{4} \mu^2 R\left(2 M_F^2+1\right) \\
& + \frac{1}{R}\left( \frac{17}{960}  +\frac{1}{48} M_F^2  \left[\log (4096)-7\right] + \frac{1}{8} M_F^2 \log (\mu  R)\right) + \mathcal{O}\left(\frac{1}{\mu}\right) \;. \nonumber
\end{align}
The $17/960 R$ term is the standard fermion Casimir energy in the massless limit.  Readers used to spectral sum calculations on $\mathbb{R}^4$ might have expected to see a $m_F^4R^3$ term, but it does not appear in the expression we showed.  The reason is that we are working in finite volume with $M_F = m_F R \ll 1$, rather than $M_F \gg 1$.  Finally, we emphasize that for any $M_F \neq 0$ there is a non-power-suppressed term which has a logarithmic dependence on the cutoff scale $\mu$.    This is typical of non-scale-invariant theories, and as a consequence in such theories it is not very useful to discuss the `finite part' in $E(\mu)$, because such terms are highly sensitive to the choice of the UV regulator. 

Now let us consider a large-$N$ gauge theory with $n_F = 2$ massive adjoint fermions, with a common mass $m_F$.  Working to leading non-trivial order in the small $M_F$ expansion, we find 
\begin{align}
\label{eq:massDeformedLargeN}
\begin{split}
&\frac{\partial}{\partial q}\log \left[1 - z_V(q) + 2 z_F(q, M) \right] =c_{-1} (1-q)^{-1}+ c_0 + c_{1} (1-q)+ c_2 (1-q)^2 \\
& \qquad+c_3 (1-q)^3 + c_{3,\ell}  (1-q)^3\log(1-q)  + \mathcal{O}\left[ (q-1)^4,  (q-1)^4 \log(1-q) \right]  \;,
\end{split}
\end{align}
where $c_{-1} = 3$, $c_0= 3/2$, $c_1 = 3/4 +M_F^2$, $c_1 = \frac{3}{8} \left(4 M_F^2+1\right)$, $c_3 =\frac{1}{24} M_F^2 \left(12 M_F^2+37-24 \log (2)\right)$, and $c_{3,\ell} = M_F^2/2$, and the notation is meant to be reminiscent of \eqref{eq:fullSeriesModular}. Note that $M_F$ shows up only in the terms which \emph{vanish} as $q \to 1$.  Moreover, the terms which have a logarithmic dependence on $\mu$ are suppressed by positive powers of $1/\mu$.  So, provided that the suppressed terms in \eqref{eq:massDeformedLargeN} can be neglected, we can leverage the arguments of Section~\ref{sec:E0} to conclude that the $\mu^0$ and $\mu^4$ coefficients continue to vanish even when we turn on small masses for the matter fields.

However, before coming to such a conclusion, we must make sure that the neglected terms do not upset the story.  As a reminder, the issue is that the way these terms enter the full spectral sum is through terms like $(1-q^n)^m$ and $(1-q^n)^m \log(1-q^n)$, which are summed over $n$.  So even though these terms are suppressed in the $q \to 1$ limit for any fixed $n$, one might worry that, because of the sum over $n$, they could contribute to the non-vanishing terms in the large $\mu$ expansion of the large-$N$ spectral sum.  More precisely, the question is whether there could be any non-vanishing contributions to the coefficients of $\mu^4$ and $\mu^0$, and whether there is any non-power-suppressed logarithmic dependence on $\mu$, which would render the coefficient of $\mu^0$ scheme-dependent.

\begin{figure}[h!]
\centering
\includegraphics[width=0.7\textwidth]{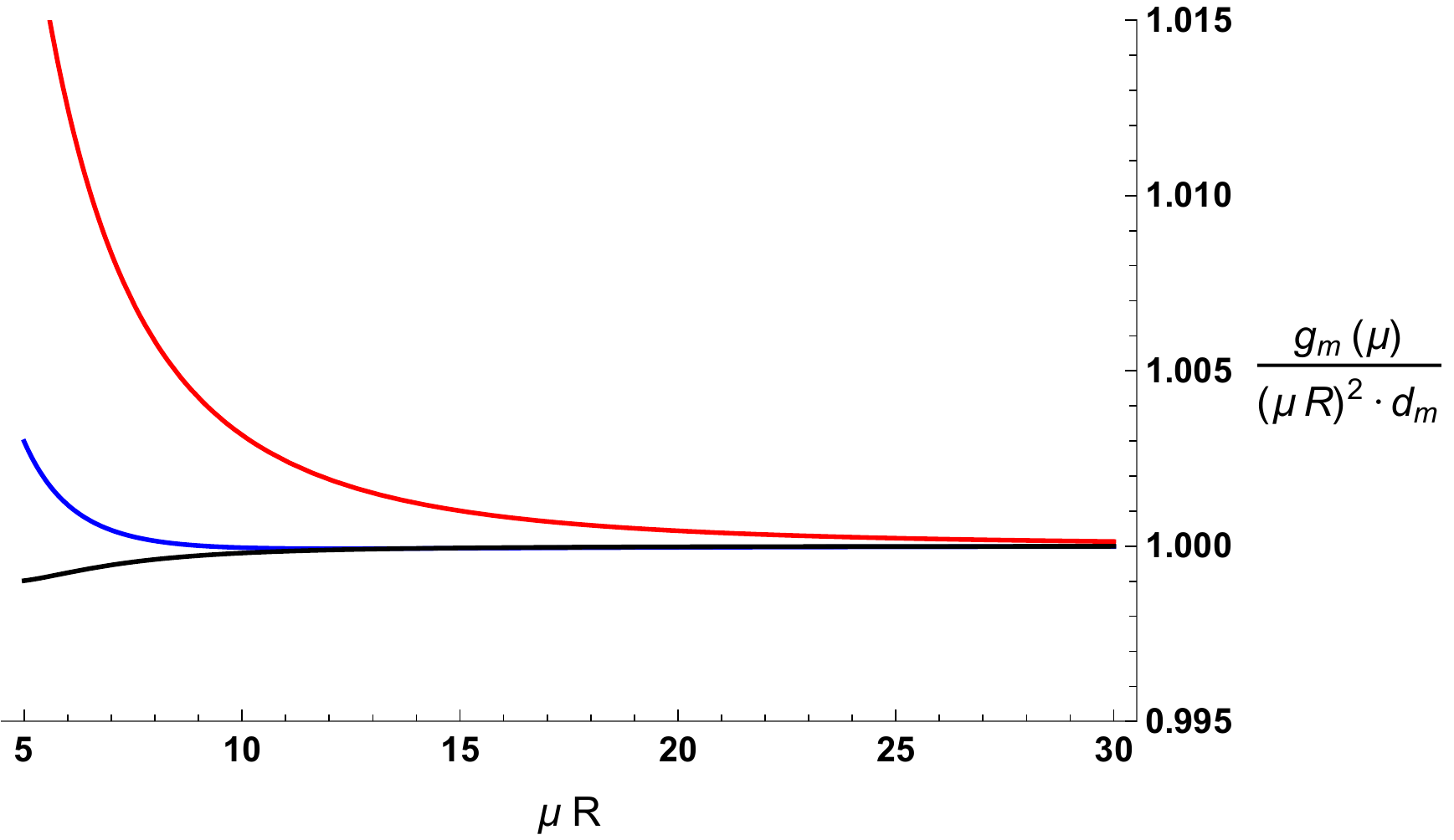}
\caption{(Color Online.) Large $\mu$ behavior of $g_{m}(\mu)$ for $m=1$ (red curve), $m=2$ (blue curve) and $m=3$ (black curve), normalized to $(\mu R)^2 d_m$, where $d_m$ is the coefficient of $\mu^2$ in \eqref{eq:gmAsymptotics}.
}
\label{fig:mu2coeffGm}
\end{figure}

Fortunately, thanks to the analysis of the functions $f_m(\mu)$ done in Section~\ref{sec:subleading}, we can already conclude that terms of the form $c_m (1-q)^{m}, m>0$ in \eqref{eq:massDeformedLargeN} do not contribute to the coefficients of $\mu^4$ or $\mu^0$ in the large $\mu$ expansion of $E(\mu)$, nor do they produce non-power-suppressed logarithmic dependence on $\mu$.  Our remaining task is to evaluate the large $\mu$ asymptotics of terms of the form
\begin{align}
g_{m}(\mu) =  \sum_{n\ge 1} \varphi(n) q^n (1-q^n)^m \log(1-q^n) \;,
\label{eq:gm_definition}
 \end{align}
with $m>0$.  To work out the asymptotics, we use the elementary identity
\begin{align}
\left(\frac{d}{dy} x^{y} \right)\bigg|_{y=0} = \log x
\label{eq:log_identity}
\end{align}
to relate $g_{m}(\mu)$ to $f_{m}(\mu)$.  In fact, the identity \eqref{eq:log_identity} implies that
\begin{align}
g_{m}(\mu) =\frac{d}{d\alpha} f_{m +\alpha}(\mu)\bigg|_{\alpha = 0} .
\end{align}
So to find the asymptotics of $g_{m}(\mu)$, we can leverage our knowledge of the asymptotics of $f_{m}(\mu)$ with $m >0$, which take the form
\begin{align}
f_m(\mu)
&= \frac{6\mu^2}{\pi^2} \frac{\psi(m+2)+\gamma_{\rm E} }{m+1} + \frac{1}{6} \cdot 0 
+ \mathcal{O}\left(\frac{1}{\mu}\right)\,.
\label{eq:fM_polylog}
\end{align}
where $\psi(z) = \Gamma'(z)/\Gamma(z)$ is the digamma function and $\gamma_{\rm E}$ is the Euler-Mascheroni  constant.   Applying \eqref{eq:log_identity} we get
\begin{align}
g_{m}(\mu) = \frac{6}{\pi^2}\left[\frac{\psi'(m+2)}{m+1}-\frac{\psi(m+2)+\gamma_{\rm E} }{(m+1)^2}\right] (\mu R)^2 + 
\mu^0 \cdot 0 + \mathcal{O}\left(\frac{1}{\mu}, \frac{1}{\mu} \log(\mu)\right) \;.
\label{eq:gmAsymptotics}
\end{align}
We note that to do this calculation, we used the analytic continuation in $m$ \eqref{eq:log_identity} of a result which was originally defined for integer $m$.  This sort of analytic continuations is a common technique for evaluating divergent quantities (the most prominent example is probably dimensional regularization of Feynman loop integrals), but when applied to a new problem it is always good to check that it gives the expected result.  In this case, we have checked that the expression for the coefficient of $\mu^2$ in $g_{m}(\mu)$ given in \eqref{eq:gmAsymptotics} agrees with a numerical evaluation of $g_m(\mu)$ for several values of $m$, as illustrated in Fig.~\ref{fig:mu2coeffGm}.  We have also done least-squares fits of \eqref{eq:gm_definition} to a polynomial in $\mu$ to estimate the value of the $\mu^0$ coefficient. The results are consistent with zero, which is the value implied by the analytic calculation.  

We are finally in a position to take stock of the situation.  The neglected terms do not affect the coefficients of $\mu^0$ and $\mu^4$ in the spectral sum. Further, they do not produce any non-power-law suppressed logarithmic dependence on $\mu$.  
At least to leading non-trivial order in the small $M_S$ and $M_F$ expansions, the sum rules \eqref{eq:sum} really do hold: both the $\mu^4$ and $\mu^0$ coefficients of the large $\mu$ expansion of the spectral sum vanish.  

\subsection{Supersymmetric matter content}
\label{sec:SUSY}

In motivating our conjecture in Section~\ref{sec:leading}, we assumed that the $(1-q)^{-3}$ terms which dominate the $q \to 1$ series expansions of the gauge fields and matter fields' single-particle partition functions --- in Eqs.~\eqref{eq:SingleLetterAsymptotics},~\eqref{eq:AsymptoticsConstraints}, and~\eqref{c1c0} --- do not cancel amongst each other. This is indeed correct when $n_F$ and $n_S$ are generic. However, for the theories with matter content corresponding to exact or softly-broken supersymmetry (SUSY), this assumption is not correct when considering spectral sums with a $(-1)^F$ twist. With adjoint matter fields, SUSY theories have matter content obeying the constraint  $n_F = p+1, n_S = 2p$ for $p \in \{0,1,2, \ldots\}$.   In supersymmetric theories, the coefficient of $(1-q)^{-3}$ in a single-letter super-multiplet vanishes, and consequently so does the coefficient of $(1-q)^{-2}$.  This is problematic in view of the computation of the spectral sums described above, because in this case the universality argument given in Sec.~\ref{sec:E0} fails, and the non-vanishing terms in the large-$\mu$ expansion of the spectral sum naively become sensitive to the subleading terms in the $q \to 1$ expansions of \eqref{eq:SingleLetterAsymptotics}.  Naively, all of the subleading terms in \eqref{eq:SingleLetterAsymptotics} are non-universal.    Nevertheless, in this section we collect some results that suggest that the large-$\mu$ asymptotics of large-$N$ spectral sums behave identically for both supersymmetric and generic gauge theories.

To support this statement, let us examine what happens to the spectral sum in large-$N$ theories on a round $S^3$ with supersymmetric matter content if we turn on mass terms for the matter fields.  The quantity we must examine is $1-z_V(q) + (p+1)z_{F}(q; M_F)- (2p) z_S(q; M_S)$, because this is the expression  that enters the $(-1)^F$-twisted single-trace partition functions.  To allow us to write explicit formulas, we can work to first non-trivial order in a small $M_S$ and $M_F$ expansion.  

As a representative example, let us consider ${\cal N} = 1$ super Yang-Mills (SYM) with $n_S = 0$, $n_F = 1$, with a mass term  $m_F \neq 0$ for the fermions, which gives a soft breaking of the supersymmetry.  To leading order in a small $M_F$ expansion near the point $q=1$, we find
\begin{align}
&1-z_v(q) + z_{f}(q; M_F) = \frac{M_F^2-\frac{3}{2}}{q-1} + \frac{1}{2} \left[M_F^2- \frac{3}{2}\right]  + \cdots \,.
\label{eq:susy1}
\end{align}
So the coefficient of $(1-q)^{-1}$, $c_{-1}$, and the coefficient of $(1-q)^0$, $c_0$, obey the relation $c_{0}/c_{-1} = -1/2$, just as was the case in theories with non-supersymmetric matter content.  The difference is that in the supersymmetric example $c_{-1}$ and $c_0$ both explicitly depend on mass parameters, while with non-supersymmetric matter content they do not depend on mass parameters.   We have checked that the conclusion that $c_{0}/c_{-1} = -1/2$ holds for any $p \ge 0$, at least to leading order in the small $M_F$ and $M_S$ expansions. 
The arguments in Sec.~\ref{sec:E0} can then be leveraged to conclude that the coefficients of $\mu^4$ and $\mu^0$ in the large $\mu$ expansion vanish, even in theories with supersymmetric matter content.  Of course, one also has to address  the loophole in the argument of Sec.~\ref{sec:E0} involving the terms that vanish as $q^n \to 1$ at fixed $n$, but the resolution of this issue is the same as in the non-supersymmetric case.

These results are consistent with two curious conclusions.  First, as anticipated in Section~\ref{sec:E0}, the calculation we used in to motivate our sum rules fails for supersymmetric theories.  Second, despite this failure, the coefficients of $\mu^0$ and $\mu^4$ follow the universal sum rule anyway.   It would be very interesting to understand why this is happening.

\subsection{Squashing of $S^3$}
\label{sec:squashing}
We now consider the effect of deforming the spatial manifold away from a round-$S^3$ geometry.  The particular deformation we will consider also breaks scale invariance.  In particular, if we regard $S^3$ as an  $S^1$ (Hopf) fibration over $S^2$, our deformation will have the  effect of changing the relative size of the base and the fiber of the Hopf fibration. This geometry preserves an $SU(2)_L \times U(1)_R$ isometry subgroup out of the original $SU(2)_L\times SU(2)_R\simeq SO(4)$ isometry group of a round $S^3$. With this choice of squashing parameter, the metric of the squashed $S^3$ can be written as
\begin{align}
ds^2=R^2\left(\omega_1^2+\omega_2^2  \right)+ \tilde{R}^2 \omega_3^2 \ ,
\label{squashed_metric}
\end{align}
where $\omega_i$ are the $SU(2)_L$-invariant 1-forms (for the unit 3-sphere). In the following we denote $\alpha := R/\tilde{R}$. The round 3-sphere corresponds to $\alpha=1$.  

To compute the letter partition functions, we need to know the eigenvalues of the scalar, spinor, and vector Laplacians on the squashed sphere.  The relevant eigenvalues, as well as their multiplicities, are summarized in Appendix \ref{app.squash_eigen}.  We do not know of a useful closed-form expression for the letter partition function for arbitrary $\alpha$.  However, working to second order in an expansion in $\Delta: = \alpha-1$, the letter partition functions take the form
\begin{align}
&z_S(q,\alpha) =\frac{q(1+q)}{(1-q)^3} +  \frac{q (1+q^2) \log(q)}{(1-q)^4} \Delta  \nonumber
  \\
&\quad+ \frac{\log (q) \Delta^2}{15 (q-1)^5}  \left[q \left(-8 q^3+36 q^2+\left(8 q^4-55 q^3+35 q^2-45 q-15\right) \log
   (q)-36 q+8\right) \right. \nonumber \\
  & \left.\qquad\qquad\qquad \qquad -8 (q-1)^5 \log (1-q)\right]+  \mathcal{O} \left[ \Delta^3 \right]\;,
  \\
&z_F(q,\alpha)=\frac{ 4 q^{\frac{3}{2}}}{(1-q)^3} +
\frac{q^{\frac{3}{2}} \left( 1+ 4q -q^2 \right) \log(q)}
{(1-q)^4} \Delta\nonumber
   \\
&\quad-\frac{\log (q) \Delta^2 }{30 (q-1)^5 \sqrt{q}} \left[q \left(-64 q^4+164 q^3-180 q^2+\left(4 q^4-91 q^3+245 q^2+55 q+75\right) q \log (q)+76 q+4\right) \right. \nonumber\\
&\left. \qquad \qquad\qquad\qquad-4 (q-1)^5 (q+1) \log (1-q)\right] +\mathcal{O} \left[\Delta^3\right]  \;,  \\
&z_V(q,\alpha)=\frac{2(3-q) q^2 }{(1-q)^3} +
\frac{4 q^2 \log (q)}{(q-1)^4} \Delta \nonumber \\
&\quad+
\frac{8 \log (q) \Delta^2}{15 (q-1)^5 q} \left[q \left(4 q^5-18 q^4+37
   q^3-37 q^2+\left(-4 q^4+20 q^3-44 q^2+35 q-25\right) q^2 \log (q)+18
   q-4\right) \right. \nonumber\\
   &\left.  \qquad \qquad\qquad\qquad 4 \left(q^2+1\right) (q-1)^5 \log (1-q)\right]   +\mathcal{O}\left[\Delta^3\right] \;.
 \label{squash_letter}
\end{align}

It is again instructive to compute the heat-kernel-regularized spectral sums for free bosons and fermions:
\begin{align}
E_S(\mu)&=6 \left(-2 \Delta^2 +\Delta -1\right) \mu^4R^3-\frac{2}{3} (5 \Delta-1) \Delta \, \mu^2 R \nonumber\\
&+\frac{1}{R}\frac{- 960 \Delta^2 \log \left(\mu R\right)-102\Delta^2+95 \Delta-15}{1800}+\mathcal{O}\left[\frac{1}{(\mu R)^2 }, \Delta^3\right] \;, \\
E_F(\mu)&= 12 \left(-2 \Delta^2+\Delta-1\right) \mu^4 R^3+\frac{1}{6} \left(26 \Delta^2-7 \Delta+3\right) \mu^2 R \nonumber\\
&+\frac{1}{R}\frac{1920 \Delta^2 \log \left(\mu R\right)-8866 \Delta^2+385 \Delta+255}{7200}+\frac{3 \Delta^2+\Delta}{24 \mu }+\mathcal{O}\left[\frac{1}{(\mu R)^2 }, \Delta^3\right]\;,
\end{align}
where $\Delta:=\alpha-1$.   The $\mu^4$ term has dependence on $\Delta$ because with our definitions, dialing $\Delta$ changes the volume of the squashed sphere, and the coefficient of $\mu^4$ depends on the space volume.  If we had defined the squashing parameter in a such a way that the squashing were volume-preserving (which is conceptually straightforward but technically inconvenient), the coefficient of $\mu^4$ would be squashing-independent.   As expected from the fact that the squashing breaks scale-invariance, the non-power-law pieces of the large-$\mu$ expansion of the spectral sums above have a  $\log(\mu)$ dependence.  This makes the $\mu$-independent terms of the large-$\mu$ expansions badly regularization dependent, already at order $\Delta^2$.

However, the situation becomes different after the matter fields couple to a confining gauge field and we take the large $N$ limit. To compute the spectral sum for confining large-$N$ gauge theories, we evaluate \eqref{c1c0} in the limit $q\to 1$, and obtain
\begin{align}
\begin{split}
&q \frac{\partial}{\partial q} \log \left[ 1-z_V(q)+n_F z_F(q)\right]  \\
&\qquad =
\frac{3 q}{1-q} + \frac{3}{2} q+
\frac{10-4 n_F-\alpha+4 n_F \alpha}{12(n_F-1) (\alpha-2)}q (1-q)  + \cdots .
\end{split}
\end{align}
The ratio of the first two terms again takes its universal form, and so we find that the coefficients of $\mu^0$ and $\mu^4$ in the large-$\mu$ expansion vanish.   The $\Delta$ dependence is present only for terms which vanish as $q^n\to 1$.  Using the same methods as in the preceding sections, it can be verified that these terms do not contribute to the coefficients of $\mu^4$ and $\mu^0$ in the large-$\mu$ expansion of the large-$N$ spectral sums.  So our conjectured sum rule continues to hold at least to leading non-trivial order in a small $\Delta$ expansion.  

\section{Discussion and future directions}
\label{sec:conclusion}

We have conjectured some universal spectral sum rules \eqref{eq2vanish} and \eqref{eq2D_Hint} for large-$N$ confined-phase gauge theories with adjoint matter in spatial boxes with the topology of $S^3$, and shown evidence that the sum rules hold in the zero-'t Hooft coupling limit even when scale-invariance is broken by mass terms for the matter fields, or by squashing the spatial manifold.  We first comment on how our work fits into a larger program motivated by \cite{Basar:2013sza,Basar:2014hda,Basar:2014jua,Basar:2015xda}, give some remarks on the relation between our spectral sums and large-$N$ vacuum energies, and conclude by outlining some potential directions for future research.

\subsection{Emergent symmetries at large $N$}

Spectral sums in quantum field theories in $d$ dimensions generally diverge as $\mu^d$, and have non-vanishing finite terms (which are often ambiguous due to logarithmic divergences).    So it is natural to wonder if there may be a symmetry-based mechanism that leads to the cancellations we have observed in our large-$N$ theories.   

A very well-known symmetry which constrains the behavior of $(-1)^F$-twisted spectral sums is supersymmetry. And indeed, one of the consequences of the analysis in~\cite{DiPietro:2014bca} is that the coefficient of $\mu^4$ divergence in $(-1)^F$-twisted spectral sums of supersymmetric QFTs must vanish, while the coefficient of $\mu^2$ is related to certain combinations of anomaly coefficients of the theory. (See also \cite{Ardehali:2015bla,Ardehali:2015hya}.) Supersymmetry also has interesting implications for the behavior of the constant term in $(-1)^F$-twisted spectral sums both in flat space \cite{Witten:1981nf} and in curved space \cite{Assel:2015nca}.

But our large-$N$ results on the vanishing of the coefficients of $\mu^4$ and $\mu^0$ coefficients hold regardless of the presence of $(-1)^F$ twist in the spectral sums, nor do they depend on whether the matter content is consistent with supersymmetry, or on the presence of SUSY-breaking mass terms.  This makes it clear that supersymmetry has nothing to do with our findings.  But then what could explain our results?   We believe that our observations can be traced to the fact that large-$N$ gauge theories have an infinite number of species of finite-mass hadrons, and to the idea that, as emphasized in several recent papers\cite{Basar:2013sza,Basar:2014hda,Basar:2015xda,Basar:2015asd}, there are reasons to expect that the distribution of hadronic states in confining large-$N$ theories is controlled by some emergent symmetries at large $N$.  Apparently, these symmetries produce interesting constraints on the spectral sums.

To see the motivation for these comments, recall that a spectral sum is in fact a combination of two conceptually different sums.  First, for each single-particle mode, there is a sum over momenta, which is divergent and must be regularized.  Second, there is a sum over the different species of single-particle modes.  In the most familiar weakly-coupled QFTs, there a finite number of particle species, and so species sums are manifestly finite.  Consequently, one can reliably estimate the behavior of the full spectral sum from the behavior of the momentum sum for each individual particle.   This is the consideration that implies that typical 4D QFTs should have spectral sums that diverge as $\mu^4$, with finite parts which scale as $M^4 \cdot \mathrm{Vol}_{\rm space}$, where $M$ is the mass of the heaviest particle.  But in theories with an infinite number of particle species, the behavior of the complete spectral sum will clearly depend on the details of the distribution of the masses of the particles, and cannot be reliably estimated from the behavior of the spectral sums for the individual particle modes, as emphasized in \cite{Basar:2014hda}. 

These comments should not be taken to suggest that any theory with a spectrum consisting of an infinite number of particle species will have interesting cancellations in its spectral sum.  For example, in supersymmetric string theories, which can of course be viewed as describing an infinite number of particles, the renormalized spectral sum (the vacuum energy)  ends up being proportional to the supersymmetry-breaking scale\cite{Rohm:1983aq}, if supersymmetry broken by the Scherk-Schwartz mechanism.  This is the same result that one would have predicted from naive field-theory considerations, so the presence of an infinite number of particle species does not lead to any unanticipated cancellations in the physical result in the systems studied in \cite{Rohm:1983aq}.  Instead, our results can be taken to be an \emph{existence proof} that there are QFTs where the sum over species \emph{can} lead to highly non-trivial cancellations.  The situation is especially intriguing because these QFTs --- confining large-$N$ gauge theories --- are of great physical interest, even before one appreciates the surprising features of their spectral sums.

The notion that infinite sums over species can lead to interesting cancellations has been anticipated in the string theory literature, for instance in \cite{Kutasov:1990sv,Kutasov:1990ua,Kutasov:1991pv,Dienes:1994np,Dienes:1994jt}, and especially in \cite{Dienes:1995pm}, which emphasized cancellations of spectral supertraces in non-supersymmetric string spectra.  And indeed, confining large-$N$ gauge theories are believed to be weakly-interacting string theories\cite{tHooft:1973jz,Witten:1979kh} with a string coupling $g_s \sim 1/N$ and string tension (in units of curvature in the dual bulk geometry) set by the size of the 't Hooft coupling.   Consequently, the $\lambda \to 0$ limit we consider is presumably related to the tensionless limit of some dual string theory.  While an explicit string-theoretic description of confining gauge theories is not known, the distribution of species produced by all known string theories is highly constrained by modular symmetries, and modular invariance of the worldsheet CFT is an essential consistency condition for a string theory.  From this perspective, it is entertaining to note that some time ago it has been argued that the modular properties of string theory partition functions imply that the distribution of high-energy states in consistent tachyon-free string theories must be such that the free energy diverges as $\mu^2$ with a naive field-theory-type UV cutoff \cite{Kutasov:1991pv}, see also \cite{Kutasov:1990sv,Dienes:1994np} and especially \cite{Dienes:1995pm}, instead of the e.g.\ $\mu^4$ divergence expected in a typical 4D theory.  This happens to be consistent with our results on the vanishing of the coefficient of $\mu^4$ in spectral sums of 4D confining large-$N$ theories.  Indeed, the vanishing of the coefficient of $\mu^4$ in our case is related to the large-$N$ modularity properties of the gauge theory partition functions uncovered in \cite{Basar:2015xda,Basar:2015asd}, as explained in Section~\ref{sec:E4}.  Our results concerning the coefficients of $\mu^0$ here and in \cite{Basar:2014hda} suggest that, first, confining gauge theories really do have emergent large-$N$ symmetries, at least in the setting we have explored, and second, that these emergent symmetries have powerful and surprising consequences.

\subsection{Vacuum energy and spectral sums}

Throughout this paper we have been careful to refer to $E(\mu)$ and $\tilde{E}(\mu)$ as regularized spectral sums, and $E_0, \tilde{E}_0$ as coefficients in a large-$\mu$ expansion.  Of course, there is also a well-known connection between these quantities and the vacuum energy of QFTs, but until now we have avoided commenting on it.  The standard vacuum energy $V$ can be written as
\begin{align}
V = \frac{1}{2}\sum_{n} (-1)^F d_n \omega_n e^{-\frac{\omega_n}{\mu}} + \textrm{(counter-terms)} \;,
\label{eq:VacuumEnergy}
\end{align}
where we used a heat-kernel UV regulator $\mu$, as in the body of the paper, and the first term is just $\tilde{E}(\mu)$.  The values of the counter-terms must be chosen to cancel the UV-divergent pieces of the spectral sum. However, the counter-terms can also have finite pieces.  Different choices of these finite pieces correspond to different choices of renormalization schemes.   Which finite counter-terms are allowed depends on the symmetries of the theory. The vacuum energy takes on a physical significance in the $\mu \to \infty$ `continuum' theory if the symmetries of the QFT are powerful enough to forbid all finite counter-terms that could shift the renormalized value of $V$. It is well known that supersymmetry is powerful enough to accomplish this in flat space, where the value of $V$ becomes an order parameter for supersymmetry breaking\cite{Witten:1981nf}.  If a field theory is coupled to a curved background spacetime, it was recently understood that superconformal symmetry is sufficient to render the supersymmetric Casimir energy scheme-independent\cite{Assel:2015nca}.

The key point, however, is that one must understand all of the symmetries of a theory before deciding whether or not $V$ is scheme-independent or not.  Our results on the behavior of  $\tilde{E}(\mu)$ and $E(\mu)$, as well as the results of \cite{Basar:2013sza,Basar:2014hda,Basar:2015xda,Basar:2015asd} strongly suggest that confining large-$N$ gauge theories have powerful emergent symmetries.  These emergent symmetries might not have a simple Lagrangian description, and are currently not well-understood.  In particular, the implications of these emergent symmetries on the possible finite counter-terms are not yet worked out, except in the simplest case of theories with massless adjoint matter compactified on a round $S^3$ discussed in \cite{Basar:2015xda,Basar:2015asd}.  So we are not yet in a position to decide on the implications of the regularized spectral sum results we found in this paper for the large-$N$ behavior of the vacuum energy.  Understanding these issues better is clearly a very interesting area for future work.

\subsection{Open issues}

Our calculations were done for compactified large-$N$ gauge theories with massive adjoint matter in the free $\lambda \to 0$ limit.  We close by listing a small selection of potentially interesting questions suggested by our results.

\begin{itemize}
\item{The most important extension is probably to understand what happens to the spectral sums at finite $\lambda$. Unfortunately, it is not clear how to efficiently evaluate the finite-$\lambda$ corrections to spectral sums.  It is conceivable that it could be done numerically following the work in \cite{Aharony:2005bq,Mussel:2009uw}.  It may also be fruitful to consider the large-$N$ behavior of $E(\mu)$ in $\mathcal{N}=4$ SYM theory, because its known integrability properties  at large $N$ (see e.g.\ \cite{Beisert:2010jr} for a review) may make the finite-$\lambda$ corrections easier to handle. }

\item{Even at $\lambda=0$, there is only a general proof of the sum rules \eqref{eq2D_Hint} and \eqref{eq2vanish} with massless adjoint matter on a round $S^3$\cite{Basar:2014hda,Basar:2015xda,Basar:2015asd}. In the non-scale-invariant cases emphasized here, we have found a highly suggestive argument supporting our conjectured sum rules, and have verified that the sum rules survive when scale invariance is slightly broken.  However, as we have explained in Section~\ref{sec:subleading}, our argument in Section~\ref{sec:E0} has a loophole.   It would be interesting to find a general argument closing this loophole, rather than checking that the loophole is harmless case by case as done in Section~\ref{sec:examples}. }

\item{The argument given in Section~\ref{sec:E0} does not apply to supersymmetric theories, yet as shown in Section~\ref{sec:SUSY} the conclusions of Section~\ref{sec:E0} hold for supersymmetric theories anyway.  Why this is happening needs to be better understood, and might conceivably shed light on how to close the loophole mentioned above.}

\item{Clearly, it would be extremely valuable to explicitly understand the nature of the emergent large-$N$ symmetries whose existence is suggested by our results. In the massless round-$S^3$ limit, these emergent symmetries turn out to be connected to a 2D description of the 4D gauge theories\cite{Basar:2015xda,Basar:2015asd}.  It is important to understand to what extent such 2D-4D relations generalize to non-scale-invariant 4D theories, and also to more explicitly understand how the symmetries which are apparent in such 2D descriptions manifest themselves directly in 4D large-$N$ gauge theories.}

\item{As we mentioned above, our results appear to have some resonance with earlier results on spectral supertraces in non-supersymmetric string theories\cite{Dienes:1995pm}.  It would be interesting to make the connection more explicit.  }

\item{In the $\lambda \to 0$ limit we worked in, gauge theories are believed to become some kind of higher-spin theories, and to have an infinite number of conserved higher-spin currents.  It has recently been observed that, at least in some contexts, higher-spin symmetries appear to be powerful enough to constrain spectral sums in ways that are highly reminiscent of our findings\cite{Giombi:2013fka,Tseytlin:2013jya,Giombi:2014iua,Giombi:2014yra,Beccaria:2014jxa,Beccaria:2014xda,Beccaria:2014zma,Beccaria:2015vaa}, leading to e.g. vanishing of the finite parts of their regularized spectral sums. It would be nice to understand the connection between our observations and such higher-spin symmetries.}

\item{We have not explored the effects of adding fundamental-representation matter fields to the gauge theories, but this would clearly be an interesting thing to do.}
\end{itemize}

We hope that some of these issues will be illuminated by future works.

\subsection*{Acknowledgements} 
We would like to thank G\"ok\c ce Ba\c sar for helpful discussions at the early stages of this project, and thank Keith Dienes for insightful comments.  AC is grateful to the participants of the INT brownbag seminar at the University of Washington for helpful feedback on a presentation of some of these results.  AC is also grateful to Sungjay Lee for comments on squashed spheres, and to Clifford Cheung, Erich Poppitz, and Mithat \"Unsal for discussions.  The research of MY is supported in part by the WPI Initiative (MEXT, Japan), by JSPS Program for Advancing Strategic International Networks to Accelerate the Circulation of Talented Researchers, by JSPS KAKENHI Grant Number 15K17634, and by Institute for Advanced Study. MY would also like to thank Aspen Center for Physics (NSF Grant No.\ PHYS-1066293), Mathematical Institute (University of Oxford) and Tsinghua Sanya International Mathematics Forum for hospitality.  The research of AC is supported in part by the U.S.\ Department of Energy under Grant DE-FG02-00ER-41132.

\appendix

\section{Single-letter partition functions on the squashed $S^3$}\label{app.squash_eigen}

In this Appendix, we present computational details for the single-letter partition functions \eqref{squash_letter}.  These partition functions can be determined from the eigenvalues of the scalar, spinor, and massless vector Laplacian operators on a squashed $S^3$.  These eigenvalues can be extracted by choosing appropriate values of $\sigma$ and $q$ in \cite{Hama:2011ea}, by choosing $\sigma=0$ in their expression\footnote{Reference \cite{Hama:2011ea} discusses 3d $\mathcal{N}=2$ supersymmetric field theories on $S^3$, but we can nevertheless extract from there the results relevant to our discussion. The field  $\sigma$ is the adjoint scalar inside the 3d $\mathcal{N}=2$ vector multiplet, which for our purposes is not present.}. In the following we denote the total spin by $j$ ($2j\in \mathbb{Z}$), and the $J_3$ component for the $SU(2)_L$ ($SU(2)_R$) spin by $m$ ($\tilde{m}$). Note that in all of the energy eigenvalues below have a multiplicity $2j+1$ coming from the  unbroken $SU(2)_L$ symmetry.

For a scalar, the eigenvalues are 
\begin{align}
E_{S}=\frac{1}{R}\sqrt{
2j(2j+2)+q^2 +\left(2m+q \right)^2 \left(\alpha^2-1 \right) }
\end{align}
with $m=j, j-1, \ldots, -j$ and multiplicity $2j+1$. Here $q$ specifies the coupling for the scalar to the curvature. The minimal coupling is $q=0$, whereas 4d conformal coupling corresponds to $q=1$. 

For a Majorana fermion, we choose $\sigma=0, q=1/2$ 
to obtain
\footnote{These are the (absolute values of) the eigenvalues of the $2\times 2$ matrix \begin{align} \frac{1}{R} \left( \begin{array}{cc} (2m+\frac{1}{2}) \alpha+\alpha^{-1} & 2(j+m+1) \\ 2(j-m) & -(2m+\frac{1}{2})\alpha \end{array} \right) \ . \end{align} }
\begin{align}
E_{F}=\frac{1}{2R}\left( \pm \alpha^{-1} + \sqrt{\left[ (4m+1) \alpha + \alpha^{-1} \right]^2 + 16 (j+m+1)(j-m) }\right)
\end{align}
with $m=j-1, j-2, \ldots, -j$. These are the $2j$ modes for $(j,j+\frac{1}{2})\oplus (j, j-\frac{1}{2})$. For $(j,j+\frac{1}{2})$, we have two extra modes:
\begin{align}
E_{F}=\frac{1}{R}\left( \left(2j+\frac{1}{2}\right) \alpha+ \alpha^{-1} \right),  \quad 
\frac{1}{R} \left(2j+\frac{3}{2}\right) \alpha \ .
\end{align}

For the massless vector field (a $U(1)$ gauge field), because the longitudinal modes cancel with the ghost contributions, we  only list the contributions from the remaining transverse modes. The transverse modes have spins $(j, j+1)\oplus (j, j-1)$ under $SU(2)_L\times SU(2)_R$. We need to be careful in the exceptional cases $j=0, 1/2$. For $j=1/2$ we have total of eight states $(\frac{1}{2}, \frac{3}{2})$, and for $j=0$ we have three states $(0,1)$.

First, we have eigenvalues
\begin{align}
E_{V}=\frac{1}{R}\left( \frac{\pm 1 + \sqrt{1+4j(j+1) \alpha^2+4m^2 \alpha^2(\alpha^2-1)}}{\alpha} \right) \ ,
\end{align}
with $m=j-1, j-2, \ldots, -(j-1)$, assuming $j\ge 1$. These are the two non-zero eigenvalues of the $3\times 3$ matrix
\begin{align}
\frac{1}{R} \left(
\begin{array}{ccc}
-2m \alpha & 2(j-m) &  0 \\
j+m+1 & 2  \alpha^{-1} & -(j-m+1) \\
0 & -2(j+m) & 2m \alpha
\end{array}
\right)
\end{align}
as found in \cite{Hama:2011ea}. These correspond to the $|m|\le j-1$ components of $(j, j+1)\oplus (j, j-1)$.

We also have, $m=\pm j, \pm (j+1)$ components of  $(j, j+1)$; these exists for $j\ge 1/2$. For $m=\pm j$ and 
\begin{align}
E_{V}=\frac{2}{R} \left(
j \alpha + \alpha^{-1}
\right)
\end{align}
For $m=\pm (j+1)$.
\begin{align}
E_{V}=\frac{1}{R} \left(2j+2
\right)\alpha
\end{align}

Finally, for $j=0$ (and hence $m=0$) we have the  spin $(0,1)$ representation, and  
\begin{align} 
E_V=\frac{1}{R} 2\alpha, \quad \frac{1}{R} 2\alpha^{-1} \ .
 \end{align}
These eigenvalues have  multiplicities $2$ and $1$, respectively, which correspond to $\tilde{m}=\pm 1, 0$ components of the spin $1$ representation.

From the eigenvalue data above, we can now write down the single-letter partition functions. For a scalar, a fermion and a gauge field, we respectively obtain
\begin{align}
z_S(q,\alpha)& = \sum_{2j\in \mathbb{Z}_{\ge 0}} (2j+1) 
\sum_{m=-j}^j q^{\sqrt{ (2j+1)^2+\left(2m+1 \right)^2 \left(\alpha^2-1 \right)} } \ , \\
z_F(q,\alpha)& = \sum_{2j\in \mathbb{Z}_{\ge 0}}  (2j+1) \, J_f(j)  \ , \\
z_V(q,\alpha)&= \sum_{2j\in \mathbb{Z}_{> 1}} (2j+1)\, J_V(j) + \sum_{2j\in \mathbb{Z}_{> 0}}  (2j+1)
\left( 2 q^{2(j \alpha+\alpha^{-1}) } + 2 q^{(2j+2) \alpha } \right) 
 \nonumber \\
& \qquad+ 2 q^{2\alpha} +q^{\frac{2}{\alpha}} \ ,
\end{align}
with
\begin{align}
J_f(j):=& \sum_{m=-j}^{j-1} q^{ \frac{1}{2} \left(-1 + \sqrt{
\left[ (4m+1) \alpha + \alpha^{-1} \right]^2 + 16 (j+m+1)(j-m) } \right) } \nonumber\\
&+\sum_{m=-j}^{j-1} q^{ \frac{1}{2} \left(1 + \sqrt{
\left[ (4m+1) \alpha + \alpha^{-1} \right]^2  + 16 (j+m+1)(j-m) } \right) } 
+ q^{ \left(2j+\frac{1}{2}\right) \alpha+ \alpha^{-1} }
+q^{ \left(2j+\frac{3}{2}\right) \alpha}  \ , \\
J_V(j):=& \sum_{m=-(j-1)}^{j-1}
q^{ \left( -1 + \sqrt{1+4j(j+1) \alpha^2+4 m^2 \alpha^2 (\alpha^2-1)} \right) / \alpha   } 
 \nonumber \\
& \qquad
+ \sum_{m=-(j-1)}^{j-1} q^{
\left(  1 + \sqrt{1+4j(j+1) \alpha^2+4 m^2 \alpha^2 (\alpha^2-1)} \right) / \alpha }  \ .
\end{align}
 Expanding the single-letter partition functions in powers of $\alpha-1$, we obtain \eqref{squash_letter}. 


\bibliographystyle{JHEP}
\bibliography{super_susy} 

\providecommand{\href}[2]{#2}\begingroup\raggedright\begin{thebibliography}{10}

\bibitem{Polyakov:1975rr}
A.~M. Polyakov, \emph{{Interaction of Goldstone Particles in Two-Dimensions.
  Applications to Ferromagnets and Massive Yang-Mills Fields}},
  \href{http://dx.doi.org/10.1016/0370-2693(75)90161-6}{\emph{Phys. Lett.} {\bf
  B59} (1975) 79--81}.

\bibitem{Polyakov:1977vm}
A.~M. Polyakov, \emph{{Hidden Symmetry of the Two-Dimensional Chiral Fields}},
  \href{http://dx.doi.org/10.1016/0370-2693(77)90707-9}{\emph{Phys. Lett.} {\bf
  B72} (1977) 224--226}.

\bibitem{Polyakov:1979gp}
A.~M. Polyakov, \emph{{String Representations and Hidden Symmetries for Gauge
  Fields}}, \href{http://dx.doi.org/10.1016/0370-2693(79)90747-0}{\emph{Phys.
  Lett.} {\bf B82} (1979) 247--250}.

\bibitem{Polyakov:1980ca}
A.~M. Polyakov, \emph{{Gauge Fields as Rings of Glue}},
  \href{http://dx.doi.org/10.1016/0550-3213(80)90507-6}{\emph{Nucl. Phys.} {\bf
  B164} (1980) 171--188}.

\bibitem{Dolan:2003uh}
L.~Dolan, C.~R. Nappi and E.~Witten, \emph{{A Relation between approaches to
  integrability in superconformal Yang-Mills theory}},
  \href{http://dx.doi.org/10.1088/1126-6708/2003/10/017}{\emph{JHEP} {\bf 10}
  (2003) 017}, [\href{http://arxiv.org/abs/hep-th/0308089}{{\tt
  hep-th/0308089}}].

\bibitem{Beisert:2010jr}
N.~Beisert et~al., \emph{{Review of AdS/CFT Integrability: An Overview}},
  \href{http://dx.doi.org/10.1007/s11005-011-0529-2}{\emph{Lett. Math. Phys.}
  {\bf 99} (2012) 3--32}, [\href{http://arxiv.org/abs/1012.3982}{{\tt
  1012.3982}}].

\bibitem{Aharony:2003sx}
O.~Aharony, J.~Marsano, S.~Minwalla, K.~Papadodimas and M.~Van~Raamsdonk,
  \emph{{The Hagedorn - deconfinement phase transition in weakly coupled large
  N gauge theories}}, {\emph{Adv.Theor.Math.Phys.} {\bf 8} (2004) 603--696},
  [\href{http://arxiv.org/abs/hep-th/0310285}{{\tt hep-th/0310285}}].

\bibitem{Basar:2014hda}
G.~Basar, A.~Cherman, D.~A. McGady and M.~Yamazaki, \emph{{Casimir energy of
  confining large $N$ gauge theories}},
  \href{http://dx.doi.org/10.1103/PhysRevLett.114.251604}{\emph{Phys. Rev.
  Lett.} {\bf 114} (2015) 251604}, [\href{http://arxiv.org/abs/1408.3120}{{\tt
  1408.3120}}].

\bibitem{Basar:2013sza}
G.~Basar, A.~Cherman, D.~Dorigoni and M.~Unsal, \emph{{Large N Volume
  Independence and an Emergent Fermionic Symmetry}},
  \href{http://dx.doi.org/10.1103/PhysRevLett.111.121601}{\emph{Phys. Rev.
  Lett. 111,} {\bf 121601} (2013) 121601},
  [\href{http://arxiv.org/abs/1306.2960}{{\tt 1306.2960}}].

\bibitem{Basar:2015xda}
G.~Basar, A.~Cherman, K.~R. Dienes and D.~A. McGady, \emph{{A 4D-2D equivalence
  for large-N Yang-Mills theory}},  \href{http://arxiv.org/abs/1507.08666}{{\tt
  1507.08666}}.

\bibitem{Basar:2015asd}
G.~Ba{\c s}ar, A.~Cherman, K.~R. Dienes and D.~A. McGady, \emph{{Modularity and
  4D-2D spectral equivalences for large-N gauge theories with adjoint matter}},
   \href{http://arxiv.org/abs/1512.07918}{{\tt 1512.07918}}.

\bibitem{Polyakov:2001af}
A.~M. Polyakov, \emph{{Gauge fields and space-time}},
  \href{http://dx.doi.org/10.1142/S0217751X02013071}{\emph{Int.J.Mod.Phys.}
  {\bf A17S1} (2002) 119--136},
  [\href{http://arxiv.org/abs/hep-th/0110196}{{\tt hep-th/0110196}}].

\bibitem{Sundborg:1999ue}
B.~Sundborg, \emph{{The Hagedorn transition, deconfinement and N=4 SYM
  theory}},
  \href{http://dx.doi.org/10.1016/S0550-3213(00)00044-4}{\emph{Nucl.Phys.} {\bf
  B573} (2000) 349--363}, [\href{http://arxiv.org/abs/hep-th/9908001}{{\tt
  hep-th/9908001}}].

\bibitem{Basar:2014jua}
G.~Basar, A.~Cherman and D.~A. McGady, \emph{{Bose-Fermi Degeneracies in Large
  $N$ Adjoint QCD}},
  \href{http://dx.doi.org/10.1007/JHEP07(2015)016}{\emph{JHEP} {\bf 07} (2015)
  016}, [\href{http://arxiv.org/abs/1409.1617}{{\tt 1409.1617}}].

\bibitem{Zuo:2015mxk}
F.~Zuo and Y.-H. Gao, \emph{{Hagedorn transition and topological entanglement
  entropy}},  \href{http://arxiv.org/abs/1511.02028}{{\tt 1511.02028}}.

\bibitem{Basar:2014mha}
G.~Basar, A.~Cherman, D.~A. McGady and M.~Yamazaki,
  \emph{{Temperature-reflection symmetry}},
  \href{http://dx.doi.org/10.1103/PhysRevD.91.106004}{\emph{Phys. Rev.} {\bf
  D91} (2015) 106004}, [\href{http://arxiv.org/abs/1406.6329}{{\tt
  1406.6329}}].

\bibitem{Guillera:2008aa}
J.~Guillera and J.~Sondow, \emph{Double integrals and infinite products for
  some classical constants via analytic continuations of lerch's transcendent},
  \href{http://dx.doi.org/10.1007/s11139-007-9102-0}{\emph{The Ramanujan
  Journal} {\bf 16} (2008) 247--270}.

\bibitem{DiPietro:2014bca}
L.~Di~Pietro and Z.~Komargodski, \emph{{Cardy Formulae for SUSY Theories in d=4
  and d=6}},  \href{http://arxiv.org/abs/1407.6061}{{\tt 1407.6061}}.

\bibitem{Ardehali:2015bla}
A.~A. Ardehali, \emph{{High-temperature asymptotics of supersymmetric partition
  functions}},  \href{http://arxiv.org/abs/1512.03376}{{\tt 1512.03376}}.

\bibitem{Ardehali:2015hya}
A.~A. Ardehali, J.~T. Liu and P.~Szepietowski, \emph{{High-Temperature
  Expansion of Supersymmetric Partition Functions}},
  \href{http://dx.doi.org/10.1007/JHEP07(2015)113}{\emph{JHEP} {\bf 07} (2015)
  113}, [\href{http://arxiv.org/abs/1502.07737}{{\tt 1502.07737}}].

\bibitem{Witten:1981nf}
E.~Witten, \emph{{Dynamical Breaking of Supersymmetry}},
  \href{http://dx.doi.org/10.1016/0550-3213(81)90006-7}{\emph{Nucl. Phys.} {\bf
  B188} (1981) 513}.

\bibitem{Assel:2015nca}
B.~Assel, D.~Cassani, L.~Di~Pietro, Z.~Komargodski, J.~Lorenzen et~al.,
  \emph{{The Casimir Energy in Curved Space and its Supersymmetric
  Counterpart}},  \href{http://arxiv.org/abs/1503.05537}{{\tt 1503.05537}}.

\bibitem{Rohm:1983aq}
R.~Rohm, \emph{{Spontaneous Supersymmetry Breaking in Supersymmetric String
  Theories}}, \href{http://dx.doi.org/10.1016/0550-3213(84)90007-5}{\emph{Nucl.
  Phys.} {\bf B237} (1984) 553}.

\bibitem{Kutasov:1990sv}
D.~Kutasov and N.~Seiberg, \emph{{Number of degrees of freedom, density of
  states and tachyons in string theory and CFT}},
  \href{http://dx.doi.org/10.1016/0550-3213(91)90426-X}{\emph{Nucl.Phys.} {\bf
  B358} (1991) 600--618}.

\bibitem{Kutasov:1990ua}
D.~Kutasov and N.~Seiberg, \emph{{Noncritical superstrings}},
  \href{http://dx.doi.org/10.1016/0370-2693(90)90233-V}{\emph{Phys.Lett.} {\bf
  B251} (1990) 67--72}.

\bibitem{Kutasov:1991pv}
D.~Kutasov, \emph{{Some properties of (non)critical strings}},
  \href{http://arxiv.org/abs/hep-th/9110041}{{\tt hep-th/9110041}}.

\bibitem{Dienes:1994np}
K.~R. Dienes, \emph{{Modular invariance, finiteness, and misaligned
  supersymmetry: New constraints on the numbers of physical string states}},
  \href{http://dx.doi.org/10.1016/0550-3213(94)90153-8}{\emph{Nucl.Phys.} {\bf
  B429} (1994) 533--588}, [\href{http://arxiv.org/abs/hep-th/9402006}{{\tt
  hep-th/9402006}}].

\bibitem{Dienes:1994jt}
K.~R. Dienes, \emph{{How strings make do without supersymmetry: An Introduction
  to misaligned supersymmetry}},
  \href{http://arxiv.org/abs/hep-th/9409114}{{\tt hep-th/9409114}}.

\bibitem{Dienes:1995pm}
K.~R. Dienes, M.~Moshe and R.~C. Myers, \emph{{String theory, misaligned
  supersymmetry, and the supertrace constraints}},
  \href{http://dx.doi.org/10.1103/PhysRevLett.74.4767}{\emph{Phys.Rev.Lett.}
  {\bf 74} (1995) 4767--4770}, [\href{http://arxiv.org/abs/hep-th/9503055}{{\tt
  hep-th/9503055}}].

\bibitem{tHooft:1973jz}
G.~'t~Hooft, \emph{{A Planar Diagram Theory for Strong Interactions}},
  \href{http://dx.doi.org/10.1016/0550-3213(74)90154-0}{\emph{Nucl.Phys.} {\bf
  B72} (1974) 461}.

\bibitem{Witten:1979kh}
E.~Witten, \emph{{Baryons in the $1/N$ Expansion}},
  \href{http://dx.doi.org/10.1016/0550-3213(79)90232-3}{\emph{Nucl. Phys.} {\bf
  B160} (1979) 57}.

\bibitem{Aharony:2005bq}
O.~Aharony, J.~Marsano, S.~Minwalla, K.~Papadodimas and M.~Van~Raamsdonk,
  \emph{{A First order deconfinement transition in large N Yang-Mills theory on
  a small S**3}},
  \href{http://dx.doi.org/10.1103/PhysRevD.71.125018}{\emph{Phys.Rev.} {\bf
  D71} (2005) 125018}, [\href{http://arxiv.org/abs/hep-th/0502149}{{\tt
  hep-th/0502149}}].

\bibitem{Mussel:2009uw}
M.~Mussel and R.~Yacoby, \emph{{The 2-loop partition function of large N gauge
  theories with adjoint matter on S**3}},
  \href{http://dx.doi.org/10.1088/1126-6708/2009/12/005}{\emph{JHEP} {\bf 0912}
  (2009) 005}, [\href{http://arxiv.org/abs/0909.0407}{{\tt 0909.0407}}].

\bibitem{Giombi:2013fka}
S.~Giombi and I.~R. Klebanov, \emph{{One Loop Tests of Higher Spin AdS/CFT}},
  \href{http://dx.doi.org/10.1007/JHEP12(2013)068}{\emph{JHEP} {\bf 12} (2013)
  068}, [\href{http://arxiv.org/abs/1308.2337}{{\tt 1308.2337}}].

\bibitem{Tseytlin:2013jya}
A.~A. Tseytlin, \emph{{On partition function and Weyl anomaly of conformal
  higher spin fields}},
  \href{http://dx.doi.org/10.1016/j.nuclphysb.2013.10.009}{\emph{Nucl. Phys.}
  {\bf B877} (2013) 598--631}, [\href{http://arxiv.org/abs/1309.0785}{{\tt
  1309.0785}}].

\bibitem{Giombi:2014iua}
S.~Giombi, I.~R. Klebanov and B.~R. Safdi, \emph{{Higher Spin
  AdS$_{d+1}$/CFT$_d$ at One Loop}},
  \href{http://dx.doi.org/10.1103/PhysRevD.89.084004}{\emph{Phys. Rev.} {\bf
  D89} (2014) 084004}, [\href{http://arxiv.org/abs/1401.0825}{{\tt
  1401.0825}}].

\bibitem{Giombi:2014yra}
S.~Giombi, I.~R. Klebanov and A.~A. Tseytlin, \emph{{Partition Functions and
  Casimir Energies in Higher Spin $AdS_{d+1}$/$CFT_d$}},
  \href{http://dx.doi.org/10.1103/PhysRevD.90.024048}{\emph{Phys.Rev.} {\bf
  D90} (2014) 024048}, [\href{http://arxiv.org/abs/1402.5396}{{\tt
  1402.5396}}].

\bibitem{Beccaria:2014jxa}
M.~Beccaria, X.~Bekaert and A.~A. Tseytlin, \emph{{Partition function of free
  conformal higher spin theory}},  \href{http://arxiv.org/abs/1406.3542}{{\tt
  1406.3542}}.

\bibitem{Beccaria:2014xda}
M.~Beccaria and A.~A. Tseytlin, \emph{{Higher spins in AdS$_{5}$ at one loop:
  vacuum energy, boundary conformal anomalies and AdS/CFT}},
  \href{http://dx.doi.org/10.1007/JHEP11(2014)114}{\emph{JHEP} {\bf 11} (2014)
  114}, [\href{http://arxiv.org/abs/1410.3273}{{\tt 1410.3273}}].

\bibitem{Beccaria:2014zma}
M.~Beccaria and A.~A. Tseytlin, \emph{{Vectorial AdS$_5$/CFT$_4$ duality for
  spin-one boundary theory}},
  \href{http://dx.doi.org/10.1088/1751-8113/47/49/492001}{\emph{J. Phys.} {\bf
  A47} (2014) 492001}, [\href{http://arxiv.org/abs/1410.4457}{{\tt
  1410.4457}}].

\bibitem{Beccaria:2015vaa}
M.~Beccaria and A.~A. Tseytlin, \emph{{On higher spin partition functions}},
  \href{http://dx.doi.org/10.1088/1751-8113/48/27/275401}{\emph{J. Phys.} {\bf
  A48} (2015) 275401}, [\href{http://arxiv.org/abs/1503.08143}{{\tt
  1503.08143}}].

\bibitem{Hama:2011ea}
N.~Hama, K.~Hosomichi and S.~Lee, \emph{{SUSY Gauge Theories on Squashed
  Three-Spheres}}, \href{http://dx.doi.org/10.1007/JHEP05(2011)014}{\emph{JHEP}
  {\bf 05} (2011) 014}, [\href{http://arxiv.org/abs/1102.4716}{{\tt
  1102.4716}}].

\end{thebibliography}\endgroup


\end{document}